# Imaging the Sub-Moiré Potential Landscape using an Atomic Single Electron Transistor

Dahlia R. Klein[1†], Uri Zondiner[1†], Amit Keren[1,2], John Birkbeck[1], Alon Inbar[1], Jiewen Xiao[1], Mariia Sidorova[1,3], Mohammed M. Al Ezzi[4,5], Liangtao Peng[6], Kenji Watanabe[7], Takashi Taniguchi[7], Shaffique Adam[4,6,8], Shahal Ilani[1*]

[1] Department of Condensed Matter Physics, Weizmann Institute of Science, Rehovot 76100, Israel
[2] Department of Physics, Technion-Israel Institute of Technology, Haifa 3200003, Israel
[3] Laboratory of Nanooptics and Plasmonics, Moscow Institute of Physics and Technology, Dolgoprudny, Moscow Oblast 141701, Russia
[4] Department of Materials Science and Engineering, National University of Singapore, 9 Engineering Drive 1, Singapore 117575
[5] John A. Paulson School of Engineering and Applied Sciences, Harvard University, Cambridge, MA 02138, USA
[6] Department of Physics, Washington University in St. Louis, St. Louis, MO 63130, USA
[7] National Institute for Materials Science, 1-1 Namiki, Tsukuba 305-0044, Japan
[8] Department of Physics and Astronomy, University of Pennsylvania, Philadelphia, PA 19104, USA
† These authors contributed equally to the work.
* Correspondence to: shahal.ilani@weizmann.ac.il

**Electrons in solids owe their properties to the periodic potential landscapes they experience. The advent of moiré lattices has revolutionized our ability to engineer such landscapes on nanometer scales, leading to numerous groundbreaking discoveries. Despite this progress, direct imaging of these electrostatic potential landscapes remains elusive. In this work, we introduce the Atomic Single Electron Transistor (SET), a novel scanning probe utilizing a single atomic defect in a van der Waals (vdW) material, which serves as an ultrasensitive, high-resolution potential imaging sensor. Built upon the quantum twisting microscope (QTM) platform[1], this probe leverages the QTM's distinctive capability to form a pristine, scannable 2D interface between vdW heterostructures. Using the Atomic SET, we present the first direct images of the electrostatic potential in one of the most canonical moiré interfaces: graphene aligned to hexagonal boron nitride[2–10]. Our results reveal that this potential exhibits an approximate $C_6$ symmetry, has minimal dependence on the carrier density, and has a substantial magnitude of ~60 mV even in the absence of carriers. Theoretically, the observed symmetry can only be explained by a delicate interplay of physical mechanisms with competing symmetries. Intriguingly, the magnitude of the measured potential significantly exceeds theoretical predictions, suggesting that current understanding may be incomplete. With a spatial resolution of 1 nm and a sensitivity to detect the potential of even a few millionths of an electron charge, the Atomic SET opens the door for ultrasensitive imaging of charge order and thermodynamic properties for a range of quantum phenomena, including various symmetry-broken phases, quantum crystals, vortex charges, and fractionalized quasiparticles.**

The behavior of electrons in a lattice is governed by the periodic potential of the host material. In naturally occurring materials, this periodicity is determined by the atomic length scale, making it extremely challenging to image the local electrostatic potential directly. Over the past decade, moiré engineering has emerged as a powerful approach to create tunable periodic potentials at the interface between two vdW materials, achieving length scales significantly larger than those in atomic lattices. A canonical example is the interface between aligned graphene and hexagonal boron nitride (G/hBN), where their lattice mismatch leads to a moiré superlattice that dramatically alters the electronic properties through its periodic potential. This moiré interface has led to numerous discoveries including the observation of the Hofstadter butterfly[3–5] and Brown-Zak oscillations[11]. More recently, when coupled with graphene multilayers, this aligned G/hBN interface has played a crucial role in stabilizing even more exotic phases, such as ferromagnetism in magic angle twisted bilayer graphene[12,13], unconventional ferroelectricity in bilayer graphene[14,15], and the fractional quantum anomalous Hall effect (FQAHE) in rhombohedral pentalayer graphene[16].

Despite the pivotal role of this moiré potential, so far it has only been indirectly inferred from transport[17,18] and optical[19,20] measurements. Unlike in transition metal dichalcogenides (TMDs), whose band edges can serve as direct markers for potential variations[21], mapping the moiré potential in a vdW interface like G/hBN requires a novel imaging technique that combines high spatial resolution with exceptional potential sensitivity. The most sensitive tool currently available for imaging electrostatic potentials is the scanning single electron transistor (SET)[22–25], which uses transport through a small island in the Coulomb blockade regime for detection. However, the spatial resolution of existing scanning SETs is constrained by their lithographic dimensions (>100 nm), preventing them from resolving potentials within a moiré unit cell. While STM experiments with a graphene sensor layer[26] have achieved intermediate spatial resolution, and recent advancements in AFM-based techniques have led to high-resolution imaging of molecules[27–29], direct visualization of moiré potentials within vdW heterostructures remains an unmet challenge.

In this work, we develop the Atomic SET, a novel scanning probe that uses a single atomic defect[30–35] as a scanning potential sensor, achieving a 1 nm spatial resolution, two orders of magnitude better than existing SETs, and a potential sensitivity of 6 μV/$\sqrt{\text{Hz}}$. This remarkable sensitivity (SI Section 8) corresponds to detecting variations of a few parts-per-million of the potential produced by a single electron charge at the distance given by the spatial resolution. Using this tool, we directly image the potential at the G/hBN moiré interface. Our measurements reveal that even at zero carrier density, the peak-to-peak potential amplitude is very large, about 63 mV, and that this potential exhibits an approximate $C_6$ symmetry around the moiré center. The observed symmetry can be theoretically accounted for by a subtle interplay of competing mechanisms, yet the magnitude of the measured potential is about twice as large as that predicted by existing models, highlighting that our understanding of even this simple moiré interface is still incomplete.

The working principle of the scanning Atomic SET is illustrated in Fig. 1a: our sensor is based on a naturally occurring defect (yellow) within a flat, few-layer-thick TMD (blue), which is stacked on a graphene source electrode (grey). The physical system that we are interested in studying (purple) is placed on a scanning QTM tip, serving as the drain electrode. At low temperature, the defect behaves as a quantum dot (QD) and the current between the two electrodes flows through it via single electron tunneling (dashed arrows). This current can be turned on and off depending on whether the QD energy level is aligned with the electrochemical potential of the electrodes, making it an extremely sensitive sensor of the local electrostatic potential felt by the defect. When the system-of-interest on the QTM tip exhibits a spatially varying electrostatic potential, $\phi(\vec{r})$ (black), scanning the tip across the fixed defect leads to variation of the local potential that gates the defect. By monitoring the tunneling current, we can directly trace this potential. The inverted imaging geometry, where the system-of-interest is on the tip and the defect remains stationary, offers us the advantage of selecting an optimal defect from a large pool of natural defects within the flat TMD layer.

We identify and characterize individual defects in a TMD layer by mapping the spatial variation of the current ($I$) between the tip and the bottom electrode at a fixed bias voltage ($V_b$), as illustrated in Fig. 1b. When the tip does not overlap a defect, the current is given only by momentum-conserving elastic[1] and inelastic (phonon-assisted)[36] tunneling. However, when the tip overlaps a defect, an additional tunneling channel opens, allowing electrons to tunnel preferentially through the defect. This results in an increased $I$ when the tip's contact area (white dashed line) coincides with a defect.

Fig. 1c shows such a measurement performed at room temperature using a trilayer $WSe_2$ barrier. The scan reveals several oblong shapes where $I$ increases with respect to the background. Evidently, each individual shape results from a separate atomic defect imaging the tip contact area. The fact that these shapes have identical spatial structure and the same magnitude of $I$ increase suggests that these images are generated by defects with identical chemical origin. In this experiment, the vdW heterostructure on our QTM tip consists of aligned G/hBN, forming a moiré superlattice[2–5]. Remarkably, even at room temperature, imaging using a defect resolves this moiré structure with high contrast. Moreover, the sharpness of the edges demonstrates that defect imaging achieves a spatial resolution of approximately 1 nm (SI Section 6). While it might be expected that transferring a moiré heterostructure onto the steep topography of a QTM tip would result in significant strains that would be magnified by the moiré superlattice[37], this measurement and a similar scan in Fig. 3a show that the moiré heterostrain is minimal, typically less than ±0.3% (SI Section 7).

Another important aspect of the defects is their energetics, which we investigate by imaging them with varying voltage bias windows at low temperature. Fig. 1d presents spatial maps of $I$ measured at $T = 0.2$ Kelvin, using a different aligned G/hBN superlattice ($\lambda_m = 14$ nm) on a QTM tip scanned over a bilayer $WSe_2$ barrier and graphene electrode. All the subsequent data presented in this work use this tip and are taken at this temperature. At high bias ($V_b = -1.25$ V, left), the defect-assisted tunneling current is small compared to the overall background arising from Fowler-Nordheim tunneling through the $WSe_2$ conduction band[38]. Nevertheless, numerous replicas of the tip contact area ("lungs"-shaped) are clearly resolved, indicating a high density of defects accessible within this large bias window (highlighted in yellow, top illustration). Reducing the bias ($V_b = -0.9$ V, center) results in fewer visible tip area replicas and a reduction of the background current. At the lowest bias ($V_b = -0.35$ V, right), we identify only a single defect within the scanning window and observe minimal residual background tunneling. Our imaging experiments primarily use these relatively rare, low-energy defects for two reasons: first, their sparse distribution ensures that only one defect overlaps the tip at a time, allowing us to produce an image of the tip with only a single defect. Second, using a low-energy defect enables us to operate near zero bias, thereby avoiding the injection of hot electrons and phonons that could potentially interfere with measuring the system's thermodynamic ground state.

In the measurements so far, the defects served only as localized pathways for current. Now we aim to harness an individual defect as a fully functional QD, capable of probing local electrostatic potential and thermodynamic quantities at a specific position. This is accomplished by adding top and bottom gates to the QTM junction (Fig. 2a, details in caption). In a prototypical QD, applying a gate voltage $V_{gate}$ results in a linear shift of the QD's electrostatic potential, producing the characteristic "Coulomb diamond" diagram shown in Fig. 2b. At zero bias, there is a specific $V_{gate}$ where the $N$ and $N+1$ charge states of the QD are degenerate, permitting current to flow. When a finite bias is applied, the conduction window expands linearly with $V_b$, creating a "diamond" shape in the $V_{gate}$-$V_b$ plane.

In our experiment, the system-of-interest is situated between the gate and the defect, making the defect a local detector of the system's chemical potential. As illustrated in Fig. 2c, the electrochemical ($V$), electrostatic ($\phi$), and chemical ($\mu$) potentials of the top and bottom graphene layers are related by $\mu = V - \phi$ for each layer. Consequently, when both layers are grounded ($V_T = V_B = 0$), their local electrostatic potentials directly track their local chemical potentials. The electrostatic potential seen by the defect is given by $\phi_D = \alpha \phi_T + (1-\alpha)\phi_B$ (SI Sections 2 and 3), where $\alpha$ is a geometric lever arm factor that depends on the relative distances between the defect and the top and bottom layers ($z_T$, $z_B$) such that $\alpha = z_B/(z_T + z_B)$. The response of $\phi_D$ to a gate voltage $V$ (top or bottom) is thus a weighted sum of the inverse electronic compressibilities $d\mu/dn$ of the top and bottom layers $\frac{d\phi_D}{dV} = \alpha \frac{d\mu_T}{dn_T}\frac{dn_T}{dV} + (1-\alpha)\frac{d\mu_B}{dn_B}\frac{dn_B}{dV}$, where $dn_T/dV$ and $dn_B/dV$ are capacitive factors determined by the junction electrostatics (SI Section 2). Consequently, the Coulomb diamond lines will curve in a way that reflects the Dirac-like $\mu(n)$ of both layers[32,34,39–42], as illustrated in Fig. 2d.

Before performing tunneling measurements through a defect, we would like to first establish the electrostatics of the QTM junction, which is controlled by three voltages: the top gate ($V_{TG}$), the bottom gate ($V_{BG}$), and the bias ($V_b$). Fig. 2e shows the measured tunneling conductance, $dI/dV \equiv dI/dV_b$, as a function of $V_{TG}$ and $V_b$ at a spatial location where the tip does not overlap a defect. The measurement reveals curved lines of reduced conductance corresponding to the charge neutrality points of the top ($n_T = 0$, purple dashed line) and bottom ($n_B = 0$, white dashed line) graphene layers. The dashed lines are fits to the electrostatic model (see SI Section 2). When the bias is below a threshold bias, $|V_b| < V_{th} \approx 65$ mV (cyan dashed lines), there is a pronounced suppression of $dI/dV$. Detailed measurements (SI Section 10) indicate that this suppression is device-specific, arising from a very large contact resistance until $V_{th}$, which sharply drops above $V_{th}$. This implies that up to $V_{th}$, the bias primarily drops across the contact rather than across the QTM junction itself. This is reflected in the fact that the $n_T, n_B = 0$ dashed lines remain vertical in this region, unaffected by bias. A similar measurement done in a perpendicular magnetic field of 5 Tesla (Fig. 2f) reveals suppressed $dI/dV$ associated with the Landau level gaps of the top ($v_T^{LL} = \pm 2, \pm 6$, purple dashed lines) and bottom ($v_B^{LL} = \pm 2$, white dashed lines) layers. With additional measurements as a function of $V_{BG}$ (SI Section 5), we fully establish how the three voltages control the densities of the two layers within the QTM junction.

We now proceed to measure the tunneling through a low-energy defect as a function of $V_{TG}$ and $V_b$ (Fig. 2g), observing an order of magnitude higher conductance. The conductance appears along a curved Coulomb diamond, where the two branches trace the conditions of the defect being in resonance with top or bottom electrode (SI Section 2). Outside of this region, $dI/dV$ is strongly suppressed, corresponding to QD states with fixed electronic charge (labeled as $N$ and $N + 1$ for generality). Within the low bias regime $|V_b| < V_{th}$, the Coulomb blockade peak is nearly independent of $V_b$ (vertical), consistent with the earlier off-defect observation that the bias drops primarily on the contact in this regime. For $|V_b| > V_{th}$, the curved edges of the Coulomb diamond reflect the compressibilities of the top and bottom layers (dotted lines are model fits). Specifically, we can identify two deflection points (arrows) that correspond to the charge neutrality points of the top and bottom layers, where the defect potential responds more strongly to gate voltage (*i.e.,* higher slope) due to the reduced compressibility of the top or bottom layer. In a similar measurement done at $B = 5$ T (Fig. 2h), the Coulomb diamond edges show steps reflecting the Landau levels gaps (dotted lines are model fits).

Having shown that a defect can measure electronic compressibility at a single point, we now turn to imaging potentials in real space. Fig. 3a shows a high-resolution spatial map of $I$ through a single low-energy defect measured at a bias of $V_b = -0.35$ V, revealing the tip's detailed moiré structure. To image the electrostatic potential, we track how the zero-bias Coulomb peak shifts as we scan the tip across the defect. Fig. 3b shows the zero-bias $dI/dV$, measured as a function of position $x$ and $V_{TG}$, scanned along the white dashed line in Fig. 3a. For each $x$ we observe a narrow Coulomb blockade peak. The top gate voltage at which this peak appears, $V_{TG}^{peak}(x)$, oscillates with the moiré periodicity along $x$, reflecting the varying electrostatic potential along the moiré superlattice, $\phi(x)$. Scanning the tip across the defect changes the defect's local potential, therefore requiring a different gate voltage to reach the Coulomb blockade resonance condition, $V_{TG}^{peak}(x) = c \cdot \phi(x) + const$. Using the proportionality factor $c$ from the junction electrostatics (SI Sections 2 and 3), we can then directly extract the moiré electrostatic potential (right y-axis, Fig. 3b).

We can now apply this technique obtain a full 2D map of the moiré potential. Fig. 3c plots the zero-bias $dI/dV$ measured as function of $x$ and $y$ (dotted box in Fig. 3a) and at several values of $V_{TG}$. At the lowest $V_{TG}$, $dI/dV$ is practically zero at every $(x, y)$ position. With increased $V_{TG}$, conductance appears along rings which repeat in the $(x, y)$ plane with the moiré periodicity. These rings correspond to equipotential lines within the moiré unit cell. Further increasing $V_{TG}$ increases the radii of the rings until they merge and disappear. From a full 3D map of $dI/dV(x, y, V_{TG})$ (SI Movie 1), we can extract $V_{TG}^{peak}(x, y)$ (SI Section 4), and from the relation above, directly determine the 2D moiré potential.

Fig. 3d plots the 2D moiré potential, $\phi(x, y)$, across a single moiré unit cell, measured at three different moiré fillings $\nu = 0 \pm 0.15$, $1.3 \pm 0.2$, and $4 \pm 1$ ($\nu = 1$ corresponds to one electron per moiré unit cell). Several key features emerge from these scans: first, the overall magnitude of the potential variation within the moiré cell is large, ranging from 55 to 63 mV. Second, this magnitude varies very minimally (~10%) with moiré filling. Additionally, the scans clearly reveal three distinct high-symmetry points: one at the center of the moiré unit cell (red point, right panel) where the potential is maximal, and two around it that are 60° apart, showing potential minima (grey and blue). The potential difference between these minima is $\Delta\phi_{minima} \approx 4$ mV, a small fraction of the overall potential scale. This suggests that while a minor $C_3$ symmetry component exists in the potential, consistent with the underlying symmetry of the moiré lattice, the overall symmetry is quite close to $C_6$.

To interpret our observations, we consider various mechanisms that have been theoretically proposed to induce potential. Following Ref. 10, we decompose the moiré Hamiltonian into three terms: an effective pseudoelectric potential ($H_0$), a pseudomagnetic field ($H_{xy}$), and a local mass term ($H_z$), each associated with a corresponding sublattice Pauli matrix (SI Section 14). The $H_0$ term arises from two primary physical mechanisms. The first is a spatially varying stacking potential, which results from the relative alignment between the graphene and hBN lattices that varies within the moiré cell. The second mechanism is the in-plane relaxation of the graphene lattice, causing stretching and compression of the C-C bonds. This leads to local variations in the Dirac point energy relative to the vacuum level, captured by the deformation potential[9,43].

Figure 4 presents the theoretically predicted stacking (Fig. 4a) and deformation pseudopotential (Fig. 4b) terms in $H_0$, as well as the pseudomagnetic field $H_{xy}$ (Fig. 4c), and label the three high-symmetry stacking sites: CB (carbon above boron), CN (carbon above nitrogen), and AA (carbon above both boron and nitrogen). Generally, a pseudoelectric potential may reflect energy changes that are not electrostatic in nature. However, theory[43] and DFT calculations[44,45] suggest that the pseudopotentials in Figs. 4a and 4b are directly accounted for by charge polarization perpendicular to the layers, and therefore manifest as real electrostatic potentials.

Graphene carriers redistribute in the plane to screen these pseudopotentials, leading to a self-consistent electrostatic potential which is measured by our detector. We capture this theoretically using self-consistent Hartree calculations (SI Section 14). Figs. 4d-f show the resulting self-consistent electrostatic potentials resulting from the terms in Figs. 4a-c. Notably, screening maintains the shape of the pseudoelectric potentials but reduces their magnitude by approximately 2.2, consistent with the predicted random-phase-approximation (RPA) dielectric constant $\epsilon = 1 + \frac{\pi}{2}\alpha \sim 2.0$, where $\alpha = \frac{e^2}{4\pi\kappa\epsilon_0 \hbar v_F}$ is graphene's fine-structure constant, $v_F$ is its Fermi velocity, and $\kappa = 3.5$ represents the hBN dielectric constant[46]. Furthermore, we find that electronic screening converts the pseudomagnetic field $H_{xy}$ into an electrostatic potential (Fig. 4f). This potential is small compared to the other two, scales linearly with $\nu$, and becomes identically zero at $\nu = 0$ (same for $H_z$, see SI Section 14). Since our experiments show only a minor $\nu$ dependence, we omit the $H_{xy}$ and $H_z$ terms in further discussions.

Both leading potential terms (Figs. 4d and 4e) exhibit a clear $C_3$ symmetry around the central CN stacking site, in contrast to the approximate $C_6$ symmetry observed in our experiments. However, examining the minima at the CB and AA stacking sites, we notice that these $C_3$ symmetries are inverted: for the first term $\phi_{AA} > \phi_{CB}$ and for the second $\phi_{CB} > \phi_{AA}$. Interestingly, the two terms compensate each other to form an almost $C_6$-symmetric total potential with a pronounced central peak (see linecut, Fig. 4g). This theoretically predicted total potential, shown in Fig. 4h, strongly resembles the experimental result, with one notable exception – the overall experimental potential scale is double the theoretical prediction. One explanation may be that theory underestimates the strains in this moiré interface[21]. However, increasing strain alone will lead to a $C_3$-symmetric potential, contrasting our observations. The large discrepancy between experiment and theory demonstrates that despite G/hBN being one of the most relevant and extensively studied moiré interfaces, there remain substantial gaps in its theoretical understanding, which has direct consequences for recent experiments that use this interface to design new states of matter (*e.g.*, FQAHE in moiré pentalayer graphene).

Finally, we show the dependence of the moiré potential on the distance from the moiré interface. Extended Data Fig. 1 presents experimental potential traces measured by two defects, located approximately 0.8 nm (D2) and 1.5 nm (D1) from the interface (SI Section 3). The data clearly show that the measured potential decays rapidly, even over these small distances. This significant drop suggests that if the detector was at a moiré distance ($h = \lambda_m$) away from the interface, it would have detected only $e^{-\frac{4\pi}{\sqrt{3}}\left(\frac{h}{\lambda_m}\right)\sqrt{\frac{\epsilon_\parallel}{\epsilon_\perp}}} \approx 10^{-4}$ of the potential, underscoring the critical importance of our Atomic SET scanning at extremely close standoff distances from the relevant physics. At the same time, these measurements also demonstrate that in thin flakes, such as pentalayer graphene, electrons can still experience a significant moiré potential (tens of mV) even on the graphene layer furthest from the moiré interface.

The Atomic SET scanning probe technique demonstrated here has a combination of features that are extremely powerful for studying a wide range of quantum materials. Its QTM geometry allows it to scan within the pristine interfaces of a variety of vdW materials. Similar to existing SETs, this technique will allow quantitative measurements of thermodynamic properties such as the electronic compressibility[24,25,47] and entropy[48], but now with two orders of magnitude improved spatial resolution – below the Fermi wavelength, magnetic length, and moiré scales of many systems. This advance extends this powerful imaging method to a much broader class of physical phenomena occurring on small scales such as Wigner crystals, topological edge states, vortex charges, symmetry-broken phases, and fractionally charged quasiparticles.

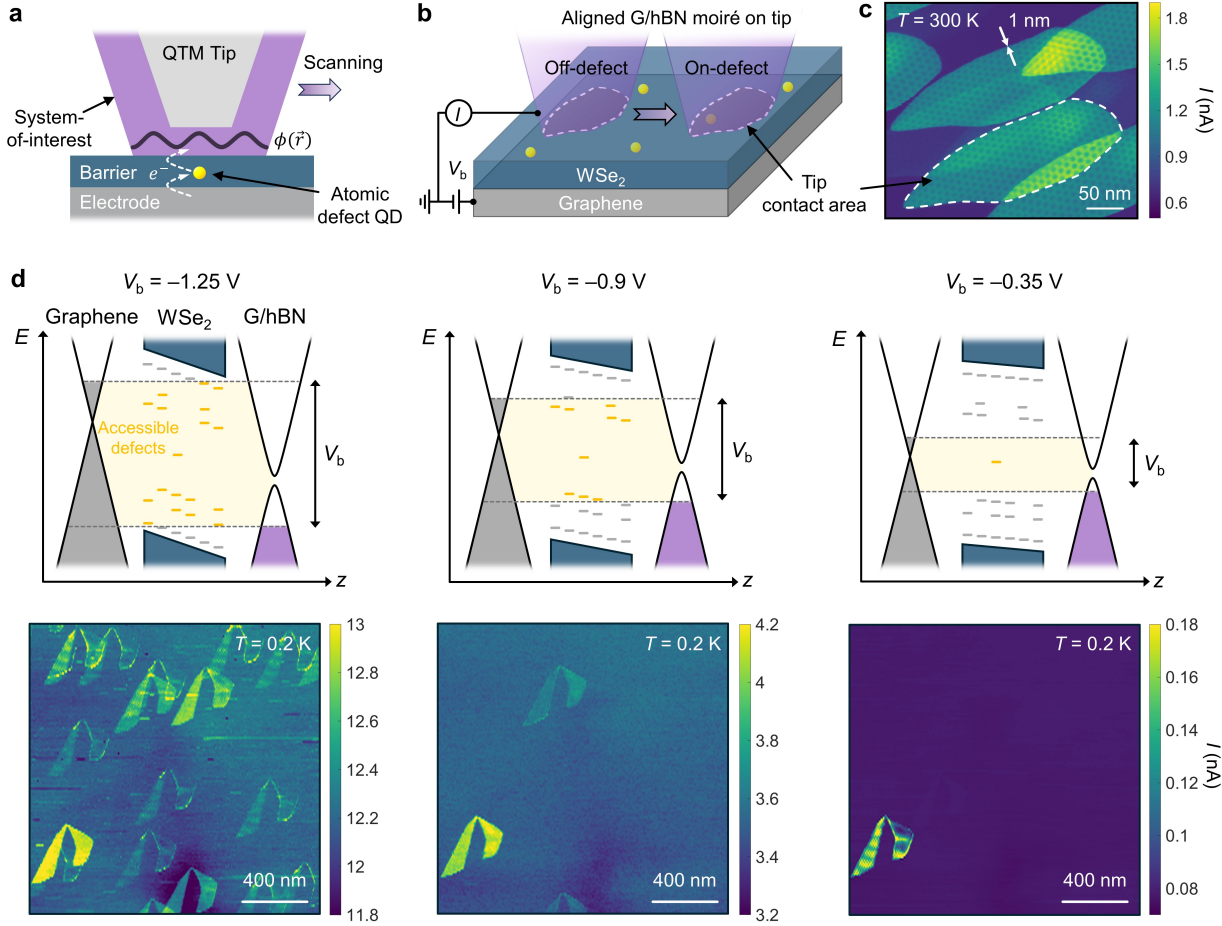

**Fig. 1: Atomic SET measurement principle and imaging of atomic defects. a.** Schematic illustration of the Atomic SET measurement principle: the system-of-interest (purple) is positioned on a QTM tip, which is scanned across a fixed atomic defect (yellow) embedded in an insulating barrier (blue) atop a graphene electrode (grey). At low temperatures, the defect functions as a quantum dot, exhibiting single electron transport (arrows). As the tip is scanned across the defect, the spatially oscillating electrostatic potential of the system-on-tip, $\phi(\vec{r})$, gates the defect. By monitoring the defect's conductance, we directly trace this potential. **b.** Illustration of an experiment to image individual atomic defects in WSe$_2$: we apply a bias voltage ($V_b$) to the bottom graphene electrode and measure the current ($I$) through the tip as it scans across the sample. When the tip does not overlap a defect ("off-defect"), the current arises from momentum-conserving tunneling processes. However, when the tip's contact area (white dashed line) overlaps a defect ("on-defect"), an additional defect-assisted tunneling pathway opens up. **c.** Defect imaging experiment performed with a QTM tip containing an aligned graphene/hBN interface (G/hBN) at its apex. We scan this tip across trilayer WSe$_2$ atop a graphene electrode at $V_b = -0.7$ V. The measured $I$ map reveals multiple replicas of the tip contact area, each produced by a different defect. Within each replica, the moiré superlattice of G/hBN is resolved, visible as periodic modulations of $I$. The sharpness of the edges (arrows) demonstrates that defect imaging has a spatial resolution of ~1 nm. **d.** Schematic energy diagrams (top) and corresponding current maps (bottom) measured at cryogenic temperatures ($T = 0.2$ K) with a different G/hBN tip at three different biases: $V_b = -1.25$ V (left), $-0.9$ V (center), and $-0.35$ V (right), all within the same scan window. The energy diagrams depict the interface along the $z$ direction, featuring a graphene source electrode (grey), defects (yellow, grey) in a bilayer WSe$_2$ barrier (blue), and a G/hBN moiré drain electrode (purple). As the magnitude of $V_b$ decreases, the range of transport-accessible defects (highlighted in yellow) is narrowed, resulting in fewer tip shape replicas in the measurement. At the lowest bias, only a single tip shape replica is seen in the entire scan window.

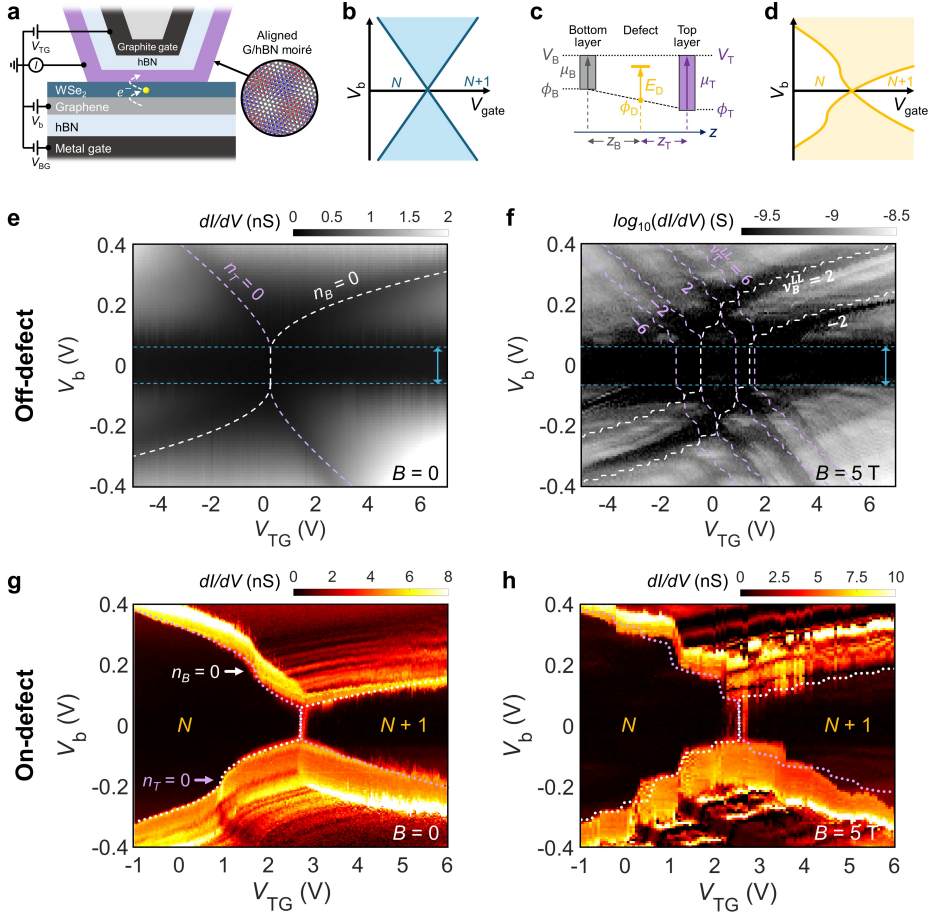

**Fig. 2: Measuring local compressibility using an atomic defect quantum dot (QD). a.** Schematic cross section of the experimental setup: the QTM tip comprises an aligned G/hBN moiré interface atop a 24 nm thick hBN layer and a graphite top gate. The sensor device contains a defect-bearing bilayer WSe$_2$ barrier, a graphene electrode, a 61 nm thick hBN layer, and a metal bottom gate. **b.** Prototypical QD "Coulomb diamond" conductance diagram as a function of gate voltage ($V_{gate}$) and source-drain bias ($V_b$). Conductance only occurs within a central region (blue), while regions outside this area exhibit Coulomb blockade, corresponding to fixed charge states (labeled $N$ and $N + 1$). **c.** Electrostatic diagram of the QTM junction, showing the relation between the electrochemical ($V$), electrostatic ($\phi$), and chemical ($\mu$) potentials of the top and bottom graphene layers given by $\mu_{T,B} = V_{T,B} - \phi_{T,B}$. The defect's electrostatic potential is given by $\phi_D = \alpha\phi_T + (1-\alpha)\phi_B$, where $\alpha = z_B/(z_T + z_B)$ and $z_T$, $z_B$ are the distances of the defect from the top and bottom layers, respectively. The defect's energy level $E_D$ is defined relative to this potential. **d.** In the experiment, the system-of-interest is located between the gate and the defect, causing screening of the gate voltage by the system's electronic compressibility. This leads to a curvature in the Coulomb diamond boundary lines, reflecting the density dependence of the chemical potential, $\mu(n)$, of the two graphene electrodes. **e.** "Off-defect" measurement (tip not overlapping a defect) of the differential conductance ($dI/dV$) vs. top gate voltage ($V_{TG}$) and bias ($V_b$) at $T = 0.2$ K and $B = 0$ T. $dI/dV$ is reduced along two curves (dashed lines), corresponding to the charge neutrality points (CNPs) of the top (purple) and bottom (white) graphene layers, indicating a reduced tunneling density of states. Additional suppression of $dI/dV$ at low bias $|V_b| < V_{th} \approx 65$ mV (horizontal cyan dashed lines) arises from nonlinear contact resistance (see text). **f.** "Off-defect" measurement at $B_\perp = 5$ T, exhibiting reduced $dI/dV$ along Landau level gaps in the top ($\nu_T^{LL} = \pm 2, \pm 6$, purple) and bottom ($\nu_B^{LL} = \pm 2$, white) graphene layers. **g.** "On-defect" measurement (QTM tip overlapping a low-energy defect) of $dI/dV$ vs. $V_{TG}$ and $V_b$ at $B = 0$ T, exhibiting a Coulomb diamond diagram with curved boundaries that separate blockaded regions with near-zero $dI/dV$ and fixed QD charge (labeled $N$ and $N + 1$) from a high $dI/dV$ region. Arrows indicate deflection points corresponding to the CNPs of the top G/hBN ($n_T = 0$, purple) and bottom graphene ($n_B = 0$, white) layers. **h.** "On-defect" measurement at $B_\perp = 5$ T, where the Coulomb diamond boundaries exhibit step-like features resulting from Landau level gaps in the two graphene electrodes. The white and purple dashed lines in **e-h** are fits to the electrostatic model (SI Sections 2 and 5).

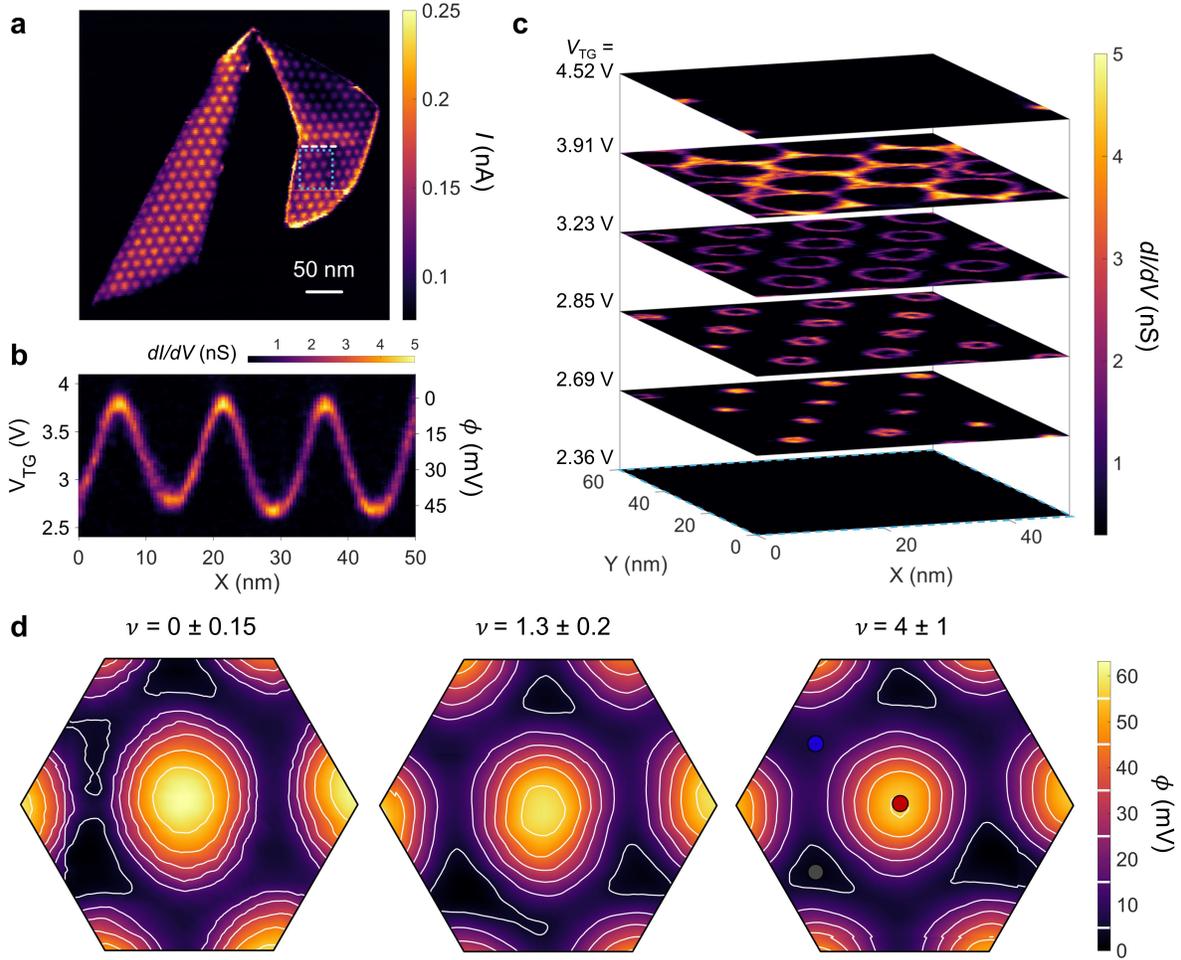

**Fig. 3: Imaging the moiré potential of aligned G/hBN using the Atomic SET. a.** High-resolution measurement of the current ($I$) vs. position ($x, y$) at $V_b = -0.35$ V and $T = 0.2$ K, revealing the shape of the G/hBN tip's contact area ("lungs" shape) and the detailed moiré structure within. **b.** Measured zero-bias $dI/dV(V_b = 0)$ as a function of the spatial coordinate $x$ and $V_{TG}$, taken at moiré filling factor $\nu = 3.2 \pm 1$. A narrow Coulomb peak is observed, whose position in gate voltage, $V_{TG}^{peak}(x)$, oscillates as a function of $x$ with the moiré periodicity. $V_{TG}^{peak}(x)$ is converted to the electrostatic potential at the moiré interface, $\phi(x)$ (right y-axis), using the junction electrostatics (SI Sections 2 and 3). This measurement cuts through the high-symmetry sites of the moiré superlattice with highest $\phi(\vec{r})$ (white dashed line in **a**). **c.** Generalization of the measurement in **b** to two spatial dimensions, showing $dI/dV$ measured vs. $x$, $y$, and $V_{TG}$, with the spatial scan taken over the area outlined by the blue dotted region in **a**. The data are plotted for a few selected slices of constant $V_{TG}$ (see SI Movie 1 for the full 3D measurement). At the lowest $V_{TG}$, $dI/dV$ is practically zero at every $(x, y)$ position. With increased $V_{TG}$, conductance appears along rings which repeat in the $(x, y)$ plane with the moiré periodicity. These rings correspond to equipotential lines within the moiré unit cell. Further increasing $V_{TG}$ increases the radii of the rings until they merge and disappear. From the full 3D map of $dI/dV(x, y, V_{TG})$, we extract $V_{TG}^{peak}(x, y)$ (SI Section 4) and from the junction electrostatics directly determine the 2D moiré potential, $\phi(x, y)$. **d.** The moiré potential, $\phi(x, y)$, of aligned G/hBN, measured at three different moiré filling factors: $\nu = 0 \pm 0.15$ (left), $\nu = 1.3 \pm 0.2$ (center), and $\nu = 4 \pm 1$ (right). These maps are obtained by averaging over several moiré sites (errors in $\nu$ arise from $V_{TG}$ gating used to meet the QD resonance condition, see SI Sections 4 and 13). We set $\phi = 0$ at the potential minimum. The center and right maps were measured with defect D1 at zero bias; the left map was measured with defect D2 at $V_b = -0.21$ V. The potential exhibits an approximate $C_6$ symmetry, changes minimally with carrier density (~10%), and has a substantial magnitude even at zero carrier density. High-symmetry stacking sites are marked by red, blue, and grey dots in the right panel.

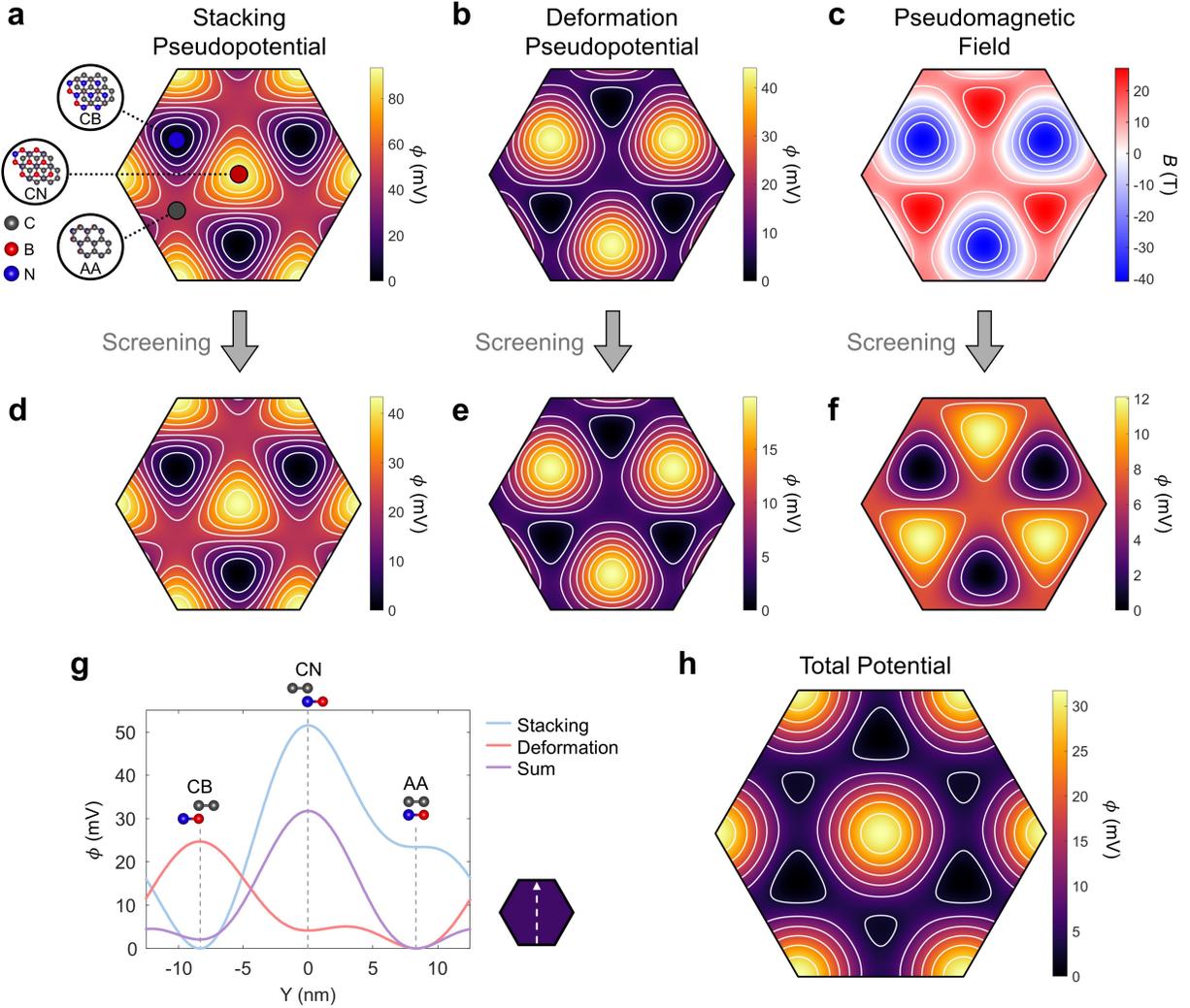

**Fig. 4: Theoretical breakdown of the various physical mechanisms contributing to the moiré potential in aligned G/hBN. a.** Stacking pseudopotential due to the relative stacking of G and hBN, which varies within the moiré unit cell. The three high-symmetry points corresponding to local CB (carbon above boron), CN (carbon above nitrogen), and AA (carbon above both boron and nitrogen) stacking are marked. **b.** Deformation pseudopotential due to atomic relaxation within the graphene layer. The terms in **a** and **b** both appear in the $H_0$ part of the Bloch Hamiltonian, corresponding to the identity matrix in the sublattice basis. **c.** Magnitude of the pseudomagnetic field, which appears in the $H_{xy}$ part of the Hamiltonian, corresponding to the $\sigma_x$ and $\sigma_y$ Pauli matrices in the sublattice basis. **d-f.** Self-consistent electrostatic potentials obtained after considering the screening by the graphene carriers, calculated by including a self-consistent Hartree potential response using a carrier density corresponding to $\nu = 4$. All terms show strong $C_3$ symmetry around the moiré center, in contrast to the $C_6$ symmetry observed in the experiments. **g.** Self-consistent stacking and deformation potentials are plotted along a linecut through the moiré center (dashed white line, bottom inset). The CB, CN, and AA high-symmetry points are labeled. Visibly, each of the two terms (blue, pink) shows a strong $C_3$ symmetry; however, due to cancelling contributions, their sum (purple) exhibits an approximate $C_6$ symmetry with only a small difference between the potential minima at the CB and AA stacking sites. **h.** Total self-consistent potential calculated for $\nu = 0$. This potential resembles the experiment in terms of the approximate $C_6$ symmetry, but its magnitude is half the one measured experimentally.

**Acknowledgments:** We thank Allan MacDonald, Hadar Steinberg, Oded Hod, Wei Cao, and Yigal Meir for fruitful discussions. This work was supported by the Israel Science Foundation under Grant No. 1621/24, the Leona M. and Harry B. Helmsley Charitable Trust grant, the Rosa and Emilio Segre Research Award, the ERC-Adg grant (QTM, no. 101097125), and the DFG funded project Number 277101999 - CRC 183 (C02). D.R.K. acknowledges support from the Zuckerman STEM Leadership Program. M.M.A.E., L.P., and S.A. acknowledge support from the Singapore National Science Foundation Investigator Award (NRFNRFI06-2020-0003) and the Singapore Ministry of Education AcRF Tier 2 grant (MOE-T2EP50220-0016). The authors thank Dr. Galit Atiya from the Materials Science and Engineering Department, Electron Microscopy Center, Technion, for her assistance with the FIB patterning.

**Author Contributions:** D.R.K., U.Z., and S.I. designed the experiment with the assistance of J.B., A.I., and J.X. in the early stages of the experiment. D.R.K. and U.Z. built the millikelvin scanning microscope. D.R.K, U.Z, A.K., and M.S. fabricated the devices and performed the experiments. D.R.K, U.Z., A.K., and S.I. analyzed the data. M.M.A.E., L.P., and S.A. wrote the theoretical model K.W. and T.T. supplied the hBN crystals. D.R.K., U.Z., and S.I. wrote the manuscript with input from other authors.

**Data availability:** The data shown in this paper are provided with the paper. Additional data that support the plots and other analysis in this work are available from the corresponding author upon request.

**Competing interests:** The authors declare no competing interests.

**Correspondence and requests for materials** should be addressed to shahal.ilani@weizmann.ac.il

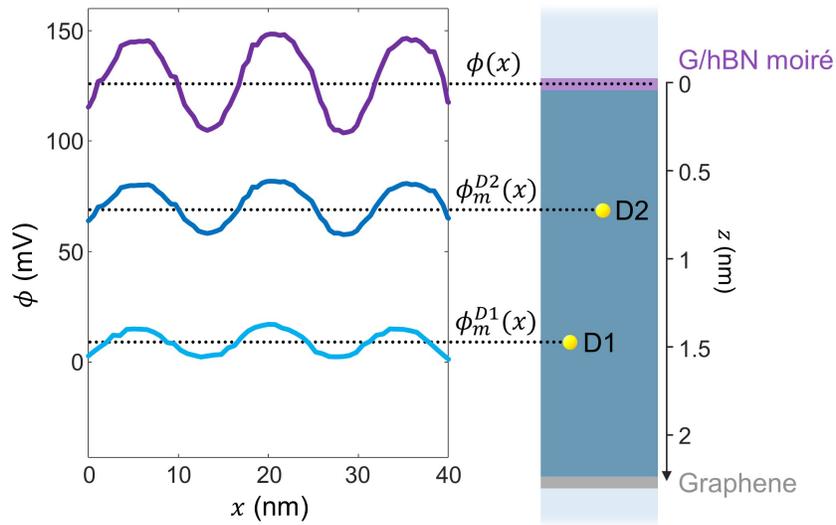

**Extended Data Fig. 1: Measured decay of the moiré potential with height.** Linecuts along high-symmetry points of the moiré potential, $\phi_m^D$, measured as a function of position, $x$, using two defects (D1 and D2). These defects are located at different heights from the G/hBN interface (1.5 nm and 0.8 nm, respectively) and measured at moiré fillings $\nu = 4 \pm 1$ and $\nu = 3.2 \pm 1$. We also plot the potential at the moiré interface ($\phi(x)$, purple) deduced using the calibrated junction electrostatics (see SI Sections 2 and 3) from the measurement with defect D2. The potentials are offset along the y-axis for clarity. These measurements demonstrate that even at such small heights, the decay of the moiré potential amplitude is substantial (~60% measured decay from 0.8 nm to 1.5 nm).

# Supplementary Information for

# Imaging the Sub-Moiré Potential Landscape using an Atomic Single Electron Transistor

Dahlia R. Klein[†], Uri Zondiner[†], Amit Keren, John Birkbeck, Alon Inbar, Jiewen Xiao, Mariia Sidorova, Mohammed M. Al Ezzi, Liangtao Peng, Kenji Watanabe, Takashi Taniguchi, Shaffique Adam, Shahal Ilani

## Table of Contents

# SI Section 1. Device fabrication

**Flat device** – The flat sensor device was fabricated using a modified dry transfer technique. Flakes of hBN and monolayer graphene (MLG, identified by Raman spectroscopy) were exfoliated onto 285 nm $SiO_2$/Si substrates. $WSe_2$ (HQ Graphene) was exfoliated onto Gel-Film X4 polydimethylsiloxane (PDMS) supported on a glass slide. A heated stamp of poly-propylene (PPC) on PDMS was used to sequentially pick up hBN, MLG, and bilayer $WSe_2$ from their substrates. Next, a flipping step was performed to invert this stack onto an $SiO_2$/Si substrate coated in a bottom layer of polyvinyl alcohol (PVA) and a top layer of poly(methyl methacrylate) (PMMA). A scribe was used to draw a circle around the stack and a droplet of water was introduced to dissolve the PVA below. The substrate was then immersed in water and a metal ring affixed to a glass slide was used to scoop out the PMMA membrane. The membrane was heated to 100ºC to allow for all water to evaporate. Finally, this stack was inverted onto a 285 nm $SiO_2$/Si substrate with pre-fabricated metallic electrodes (5/30 nm Ta/Pt) including a bottom gate (BG) and melted at 180ºC. The residual PMMA was removed with acetone. In order to electrically connect the graphene to a metal electrode, graphite was exfoliated onto a PMMA/PDMS stamp and released onto the device. After a second acetone soak, the device was annealed in vacuum for 5 hours at 300ºC to reduce bubbles and remove any remaining polymer residues. The chip was then mounted and wire-bonded to a PCB and cleaned a final time using AFM brooming performed in contact mode. An optical microscope image of the final device is shown in Fig. S1.

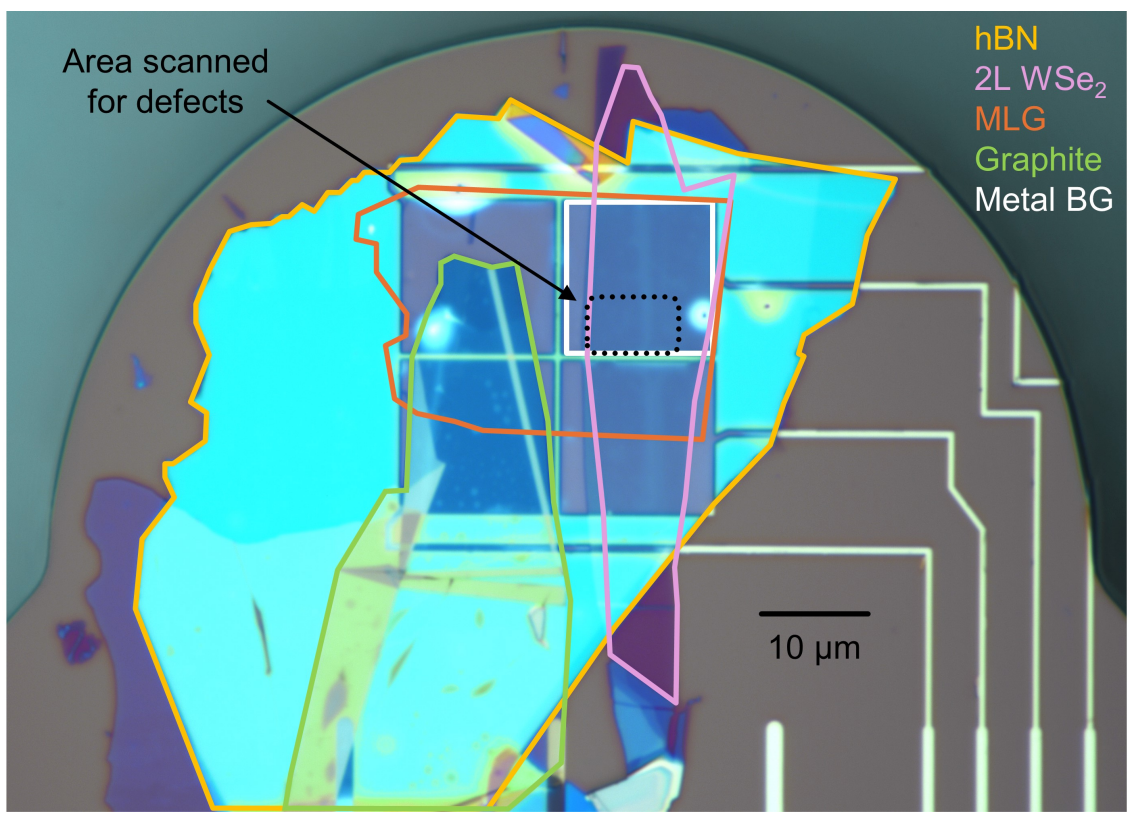

**Fig. S1.** Optical microscope image of the final sensor device with the primary scan window outlined.

**Aligned G/hBN on QTM tip** – The aligned G/hBN device was fabricated on an AFM cantilever with piezoresistive self-sensing with dimensions 120x400 µm and a spring constant of ~5 N/m. This cantilever[1], operating at room and cryogenic temperatures, allows for electrical readout of the tip deflection. This deflection readout is then used in a PID loop for feedback to the z-scanner to maintain constant deflection during scanning. This ensures that both the force and the tip contact area remain fixed during all electrical measurements. We first used FIB to deposit a small Pt pyramid, similar to previous work[2], and two Pt pads that connect to electrodes. We also used FIB to cut the metallic region between the two pads. We mounted and wire-bonded the cantilever to a PCB prior to vdW assembly. Next, we constructed the vdW heterostructure through successive layers of polymer membrane transfer and acetone washing between steps: a graphite gate and a thin hBN layer were sequentially transferred using PPC membranes. Then, we used a PPC/PDMS stamp to pick up another thin hBN layer and MLG from SiO$_2$/Si wafers, intentionally aligning the long edges of the two crystals and making sure that the hBN layer covered the entire MLG area. We confirmed alignment of graphene and hBN by the observation of broadening of the 2D peak (FWHM = 37 cm$^{-1}$) in graphene using Raman spectroscopy[3]. We then dropped this moiré heterostructure onto a PMMA/PVA/SiO$_2$/Si substrate, thereby inverting the stack. We used the same water-assisted procedure as described above to transfer this stack onto the tip. After removing the PMMA with acetone, we performed a final step of transferring graphite on a PPC membrane to bridge the MLG to a Pt contact. An optical image of the completed tip device is shown in Fig. S2.

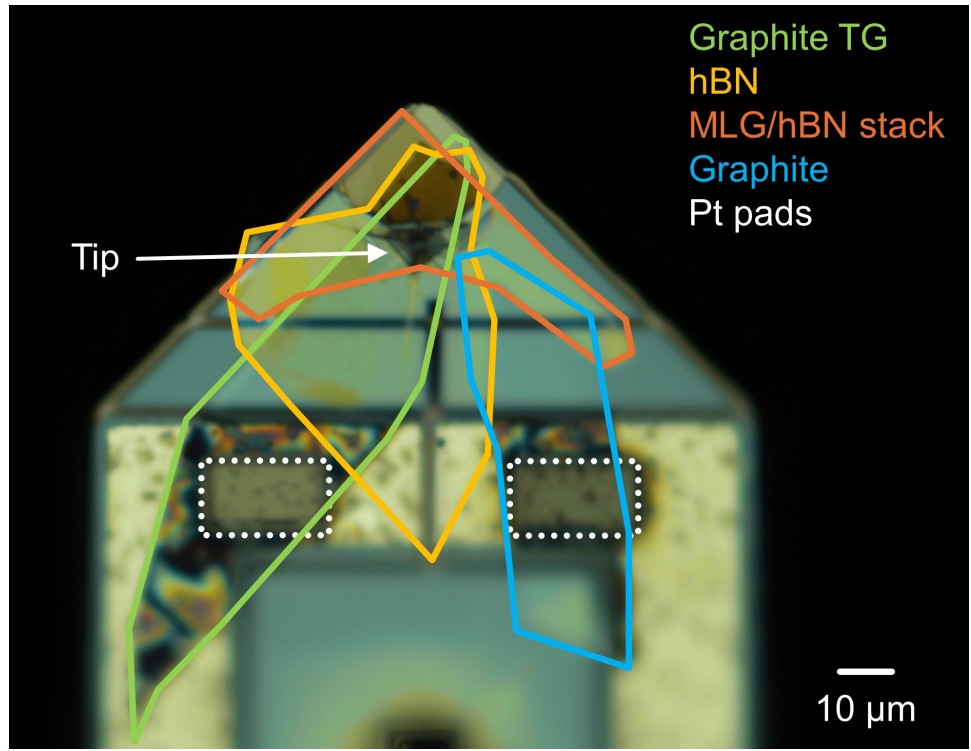

**Fig. S2.** Optical microscope image of the final device on the tip.

# SI Section 2. Electrostatic model of the QTM junction with an atomic defect and extracting the moiré potential

A unique feature of the scanning Atomic SET is that the scanning sensor, a single atomic defect, is embedded within an otherwise translationally invariant host material. Usually, one of the challenging aspects of scanning probes is that the presence of the scanning tip alters the local electrostatic environment at the point of measurement. This occurs because tips are typically made from materials with different work functions than the sample that they probe. Moreover, their geometries near the apex are often irregular, and even if they form a perfectly spherical tip, this still introduces gradients in the local gating. As a result, the potential landscape imposed by the tip on the sample can be large and unpredictable. The advantage of the QTM junction lies in its well-defined geometry. When the QTM tip comes into contact with the flat substrate, it forms a pristine, flat interface similar to the interfaces obtained within traditional stacked vdW devices. As the QTM tip scans across a TMD layer containing a single atomic defect, from the perspective of the tip, the host TMD material remains translationally invariant and only the defect appears to move. This well-defined junction geometry allows for precise control and understanding of the electrostatics. Even when the layers on the tip gate those on the bottom device (or vice versa), the planar geometry ensures that these effects can be accurately captured. In this section, we first present the electrostatic equations for the QTM junction of our experiment without an overlapping defect. These equations form the foundation for the subsequent derivation of how the electrostatics influence measurements involving a defect.

To calculate the electrostatic of the junction, we begin by using a model[4] that captures the effects of the applied voltages on the chemical and electrostatic potentials of the various layers in the junction. We will later expand this model to also describe the measurements using an atomic defect QD (Fig. S3).

We start by considering a translationally invariant system, which captures the $q = 0$ components of potential in our experiment. The model is defined by the following equations:

$$en_{TG} = C_{TG}(\phi_T - V_{TG}) \tag{S2.1}$$

$$en_T = C_{TG}(V_{TG} - \phi_T) + C_{TB}(\phi_B - \phi_T) \tag{S2.2}$$

$$en_B = C_{TB}(\phi_T - \phi_B) + C_{BG}(V_{BG} - \phi_B) \tag{S2.3}$$

$$en_{BG} = C_{BG}(\phi_B - V_{BG}) \tag{S2.4}$$

where $n_i$ describe the carrier densities of the top gate ($n_{TG}$), top electrode ($n_T$, here graphene/hBN moiré), bottom electrode ($n_B$, here graphene), and bottom gate ($n_{BG}$). $V_i$ describe the electrochemical potentials of these four plates (same subscripts) set by the external voltage sources and $\phi_i$ describe their electrostatic potentials. Since the top and bottom gates are made from graphite/metal with high density of states, in the equations above we assume that $\phi_{TG} = V_{TG}$ and $\phi_{BG} = V_{BG}$. The fixed areal geometric capacitances are defined as $C_{TG}$ (between top gate and top electrode), $C_{TB}$ (between top and bottom electrodes), and $C_{BG}$ (between bottom electrode and bottom gate). Fig. S3a labels these geometric capacitances on a schematic cross section of the junction and provides the insulating barrier layer thicknesses. The quantum capacitances of the top and bottom graphene layers are represented in this model as the capacitors $C_i^Q = e(dn_i/d\mu_i)$, shown in Fig. S3b.

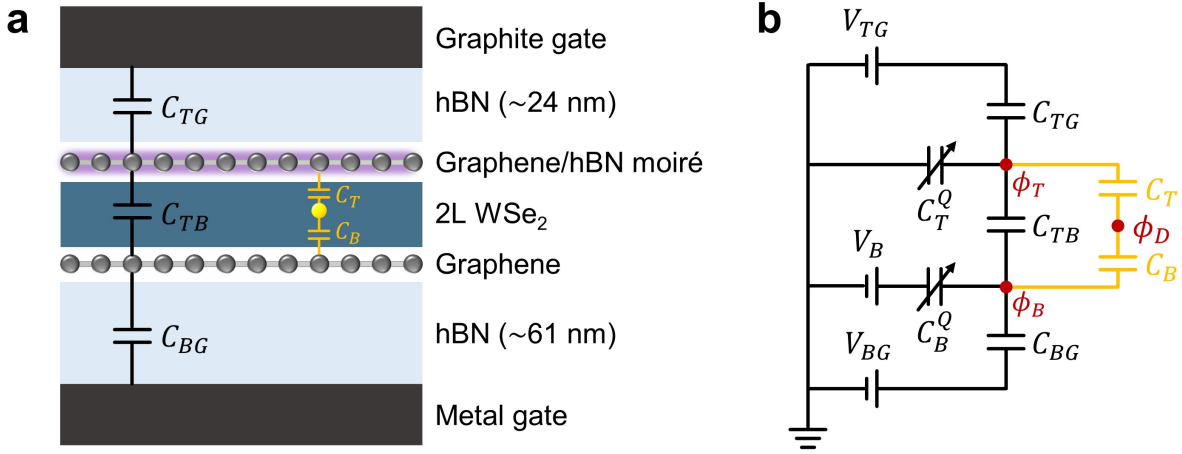

**Fig. S3: Setup of electrostatic model. a.** Schematic of the cross section of the experiment. The atomic defect is indicated as a yellow dot in the bilayer WSe₂ layer and couples to the top and bottom electrodes through relative capacitances $C_T$ and $C_B$, respectively. **b.** Equivalent circuit for a linearized model used to simulate the electrostatics of our experimental system.

For both the top and bottom graphene layers, the electrochemical ($V_i$), electrostatic ($\phi_i$), and chemical ($\mu_i$) potentials are connected via:

$$V_T - \phi_T = \mu_T(n_T) \tag{S2.5}$$

$$V_B - \phi_B = \mu_B(n_B) \tag{S2.6}$$

In our experiment, we connect the current preamplifier (not drawn) to the top graphene layer, effectively keeping this layer grounded, namely $V_T = 0$.

Using the equations above, we can fully solve how the potentials of various layers of the QTM junction depend on the applied voltages. These potentials, in turn, determine the potential at the QD location, $\phi_D$, through relative capacitances $C_T \propto \frac{1}{z_T}$ and $C_B \propto \frac{1}{z_B}$ to the top and bottom electrodes, respectively, where $z_T$ and $z_B$ are the relative distances between the QD and the two layers:

$$\phi_D = \frac{\phi_T C_T + \phi_B C_B}{C_T + C_B} \tag{S2.7}$$

We assume that the dot does not affect the charging of the layers, *i.e.*, $C_T$ and $C_B$ are much smaller than all other capacitances.

In addition to the spatially homogeneous ($q = 0$) component of the potential discussed above, the top layer also has a moiré potential, $\phi_m(x, y)$. This potential decays as a function of the distance from the top layer, $z$, as:

$$\phi_m^z(x, y) = \phi_m(x, y) F(z) \tag{S2.8}$$

where the function $F(z)$, describing the vertical decay, is analyzed in detail in SI Section S3. Specifically, we define the moiré potential at the moiré interface itself ($z = 0$) to be $\phi_m(x, y)$ and at the position of the defect ($z = h_D$) to be $\phi_m^D(x, y)$. This potential adds to the spatially homogeneous component to give the following potential at the defect position:

$$\phi_D(\{V_i\}, x, y) = \phi_m^D(x, y) + \frac{\phi_T(\{V_i\})C_T + \phi_B(\{V_i\})C_B}{C_T + C_B} \tag{S2.9}$$

We note that this equation is valid when $\phi_m(x, y)$ is slowly varying as a function of $n_T$, as is the case in our experiment.

At equilibrium and as a function of gating, a QD transitions between integer charging states (which in our experiment we refer to as $N$ and $N + 1$ for generality). At a finite bias, this transition can become resonant whenever the electrochemical potential of the top or bottom layer (effectively the source and drain electrodes of the QD) is equal to the electrochemical potential of this transition, $V_D(N \to N + 1)$. Defining $E_D \equiv V_D(N \to N + 1) - \phi_D$, these conditions become:

$$E_D + \phi_D = V_T \tag{S2.10}$$

$$E_D + \phi_D = V_B \tag{S2.11}$$

At zero bias ($V_T = V_B = 0$), the dot is resonant with both electrodes when $E_D = -\phi_D$. Substituting this into equation (S2.9), we obtain:

$$\phi_m^D(x, y) = -E_D - \frac{\phi_T(\{V_i^{peak}(x,y)\})C_T + \phi_B(\{V_i^{peak}(x,y)\})C_B}{C_T + C_B} \tag{S2.12}$$

where $\{V_i^{peak}\}$ are a set of voltages that satisfy the resonance condition at position $(x, y)$.

To derive the relation between $V_{TG}^{peak}(x, y)$ and $\phi(x, y)$ described in the main text, we linearize equation S2.12 with respect to $V_{TG}$ around the gating conditions $\{V_i^{peak}(x_0, y_0)\}$, where $\phi_D(\{V_i^{peak}\}, x, y)$ is equal to a constant. This gives us the following equation:

$$\phi_m^D(x, y) = -\frac{\partial}{\partial V_{TG}} \left[ \frac{\phi_T(\{V_i^{peak}(x_0, y_0)\}, x, y)C_T}{C_T + C_B} + \frac{\phi_B(\{V_i^{peak}(x_0, y_0)\}, x, y)C_B}{C_T + C_B} \right] V_{TG}^{peak}(x, y) + const. \tag{S2.13}$$

Including the decay factor $F(z = h_D)$, derived in the next section, we obtain:

$$V_{TG}^{peak}(x, y) = c \cdot \phi(x, y) + const. \tag{S2.14}$$

$$\frac{1}{c} = -\frac{1}{F(h_D)} \frac{\partial \phi_D}{\partial V_{TG}} = -\frac{1}{F(h_D)} \frac{1}{C_T + C_B} \left[ \frac{d\mu_T}{dn_T} \frac{dn_T}{dV_{TG}} C_T + \frac{d\mu_B}{dn_B} \frac{dn_B}{dV_{TG}} C_B \right] \tag{S2.15}$$

# SI Section 3. Height dependence of the moiré potential

In this section we derive how a spatially varying potential decays with height. Since each defect measures the moiré potential at a height of approximately 1 nm from the moiré interface, understanding this decay is essential for determining the potential at the interface from our measurements. More broadly, understanding this decay is also important for experiments that study electrons confined to a certain height above a moiré interface. One example is rhombohedral pentalayer graphene aligned to hBN[5], which does not show the fractional quantum anomalous Hall effect when the electrons are at the moiré interface, but exhibits this effect on the opposite surface of the rhombohedral graphene, further away from the moiré interface.

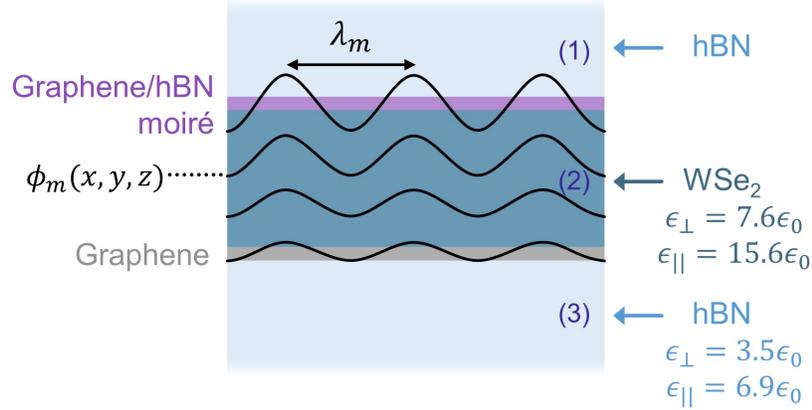

**Fig. S4: Modeling the decay of the moiré potential.** Schematic of the decay of the spatially varying moiré potential $\phi_m(x, y, z)$ as a function of height and dielectric environment. The moiré potential is screened by the layers in the QTM junction: hBN (regions 1 and 3), bilayer WSe$_2$ (region 2), and bottom graphene layer (grey).

We start with the anisotropic Laplace's equation that accounts for the dielectric environments in the QTM junction (Fig. S4):

$$\vec{D} = \epsilon(z)\vec{E} = \epsilon(z)\vec{\nabla}\phi \tag{S3.1}$$

Within the WSe$_2$ and bottom hBN layers (regions (2) and (3) in Fig. S4, respectively), Laplace's equation satisfies:

$$\epsilon_\parallel \nabla_\parallel^2 \phi = -\epsilon_\perp \partial_z^2 \phi \tag{S3.2}$$

At the bottom graphene layer ($z = d$), the jump in $D_z$ is equal to the graphene surface charge density:

$$\Delta D_z = \epsilon_\perp^{(2)} \partial_z \phi^{(2)} - \epsilon_\perp^{(3)} \partial_z \phi^{(3)} = \sigma^{graphene} = -\beta k \phi \tag{S3.3}$$

where the index $(i)$ labels the regions as in Fig. S4, $\beta$ is a parameter related to the bottom graphene screening, and $k$ is the relevant wavevector given by the reciprocal vector of the moiré lattice or its integer multiples. We now consider several variations of graphene screening. The extreme cases treat graphene as a perfect metal ($\beta \to \infty$) or as a perfect insulator ($\beta = 0$). A more accurate case considers graphene at charge neutrality, where interband electron-hole transitions generate partial screening[6] with $\beta = 3.44$, which we use in our calculation. We take the ansatz $\phi = A_1 e^{\alpha^{(i)} z} e^{i\mathbf{k}\cdot\mathbf{r}} + B_1 e^{-\alpha^{(i)} z} e^{i\mathbf{k}\cdot\mathbf{r}}$ in each region of space, where $\alpha^{(i)} = \sqrt{\frac{\epsilon_\parallel^{(i)}}{\epsilon_\perp^{(i)}}} |\mathbf{k}|$ for WSe$_2$ and hBN.

We use $\epsilon_\perp = 3.5\epsilon_0$ and $\epsilon_\parallel = 6.9\epsilon_0$[7] for hBN, and $\epsilon_\perp = 7.6\epsilon_0$ and $\epsilon_\parallel = 15.7\epsilon_0$ for bilayer WSe$_2$[7]. Using our experimental fit to the capacitance of the WSe$_2$ layer (see SI Section 5), we obtain a thickness of $d = 2.3$ nm. In Fig. S5, we plot the decay function $F(z)$ as a function of height $z$ for $\beta = 3.44$ (blue), the limiting conditions where $\beta = 0$ (green) and $\beta \to \infty$ (orange), and the translationally invariant ($q = 0$) case discussed in the previous section ($\lambda \to \infty$, black).

We estimate the heights of the two defects used in our potential maps from their relative capacitive couplings to the top and bottom layers. We then use the decay function to determine their corresponding decay factor $\phi_m^D/\phi_m$. The defect further away from the G/hBN moiré interface (which we call D1) has a decay factor of 0.31 and was used to measure the potential maps at $\nu = 1.3 \pm 0.2$ and $\nu = 4 \pm 1$. The closer defect (which we call D2) has a decay factor of 0.54 and was used to measure the potential map at $\nu = 0 \pm 0.15$. These defect positions are marked by grey dashed lines in Fig. S5. The maps presented in Fig. 3d of the main text show the moiré potential at the interface, $\phi_m(x, y)$, after taking into account the relevant decay factors. In the main text, all stated potentials $\phi$ refer to $\phi_m$ unless otherwise specified.

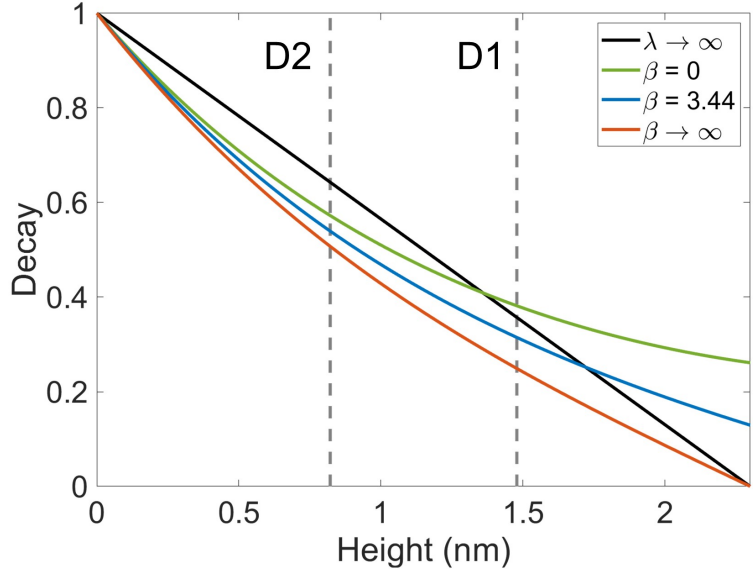

**Fig. S5.** Decay function vs. height for several limits of bottom graphene screening, parameterized by a factor $\beta$ (green, blue, and orange, see legend). The case of spatially homogenous ($q = 0$) potential ($\lambda \to \infty$) is shown in black. The vertical positions of defects D1 and D2, determined by independent measurements of the ratios of their capacitances to the top and bottom layers, are marked with grey dashed lines.

# SI Section 4. Extracting the moiré potential from measured conductance maps

In the following, we describe our procedure to convert maps of conductance vs. gate voltage and position into electrostatic potential maps. At each position $(x, y)$ we fit a Gaussian function to the $dI/dV$ measured as a function of $V_{TG}$ at zero DC bias and constant $V_{BG}$. This fit allows us to extract a 2D map of the gate voltage of the conductance peak, $V_{TG}^{peak}(x, y)$, from a 3D dataset $dI/dV(x, y, V_{TG})$. In Fig. S6a we plot this map obtained using defect D1 and $V_{BG} = -4.5$ V. Next, we calculate $\phi_T(\{V_i^{peak}\})$ and $\phi_B(\{V_i^{peak}\})$ for the translationally invariant case using equations S2.1-4 with $V_T = V_B = 0$. From equation S2.12, this gives us a map of $\phi_m^D(x, y)$, up to an additive constant proportional to $E_D$, which we define to be 0 such that $\phi_m^D = 0$ at its minimum. We then use the decay function $F(z)$ described in SI Section S3 to convert this map to $\phi_m(x, y)$ (Fig. S6b). In this figure, we can see that in addition to the potential modulation that follows the moiré periodicity, there is also a smooth, longer-range disorder potential (visible, *e.g.*, by looking at how the height of the potential peaks vary between moiré sites). In Fig. S6c, we remove this slowly varying background potential by using a biharmonic spline interpolation, enforcing that the potential peaks in different moiré cells have the same value. Finally, we obtain the map in Fig. S6d by averaging over several moiré unit cells. By performing this procedure with different defects and under different gating conditions, we can compare electrostatic potential maps $\phi_m(x, y)$ for different fillings $\nu$ of the moiré superlattice.

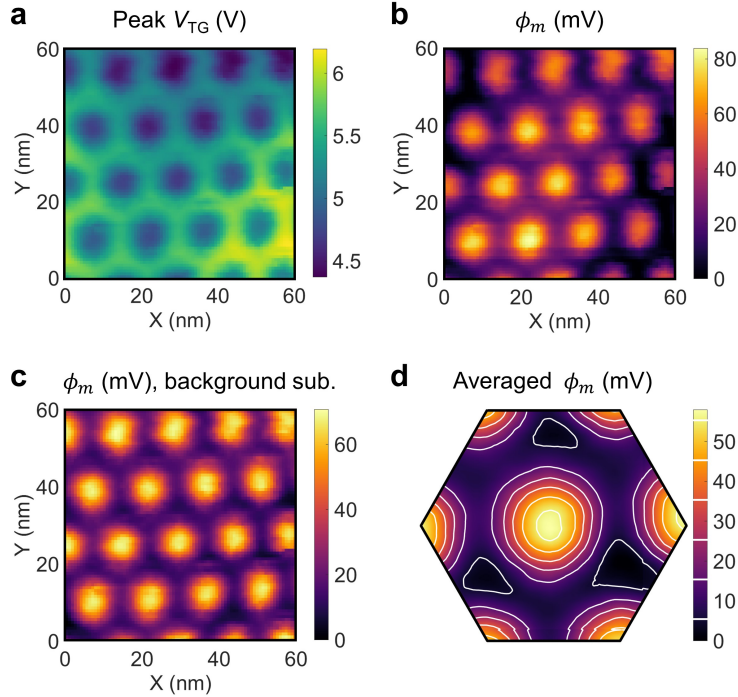

**Fig. S6: Conversion of a conductance map vs. gate voltage and position to an electrostatic potential map. a.** Map of the top gate voltage value $V_{TG}^{peak}(x, y)$ where the conductance is maximal. **b.** Map of the electrostatic potential $\phi_m(x, y)$ on the moiré graphene layer determined from the measurement in **a** using the electrostatic model described in SI Section 2. **c.** Map from **b** after removal of long-range background disorder potential. **d.** Map of the moiré potential around a moiré unit cell obtained by averaging over several unit cells in the map in **c**. This map appears in the main text as the right panel of Fig. 3d.

## SI Section 5. Calibration of geometric capacitances

We calibrated the geometric capacitance of the bottom gate using Landau level measurements in a perpendicular magnetic field. Figs. S7a and S7b plot the measured off-defect junction conductance $dI/dV$ and transconductance $dI/dV_{BG}$, respectively, as a function of $V_{BG}$ and $V_{TG}$ at an applied magnetic field of 5 Tesla. The measurements were performed with zero DC bias, and with AC excitations of 50 mV on the sample and 30 mV on the bottom gate. The vertical lines of decreased $dI/dV$ correspond to Landau level gaps in the bottom graphene electrode *outside* of the tip contact area (*i.e.*, in the region unaffected by the top gate voltage). The white dashed lines in the figure correspond to the gaps at $\nu_B = \pm 2$. Using the Landau level quantization condition $\Delta n = \left(\frac{eB}{h}\right)\Delta\nu$ and the relation $\Delta n = C\Delta V = \left(\frac{\varepsilon}{ed}\right)\Delta V$, we can extract the areal capacitance $C_{BG}$. Taking $\varepsilon_{hBN} = 3.5\varepsilon_0$, we find that $d_{BG} = 61$ nm. We also identify that the center of these Landau level gaps occurs at a finite $V_{BG} = 0.55$ V, reflecting a small intrinsic doping in the sample, and account for this in an effective $V_{BG}$ used in our electrostatic model calculations.

The additional capacitances were calibrated through junction tunneling without a defect at 5 Tesla with fixed $V_{BG} = 0$. Fig. 2f in the main text shows the off-defect $dI/dV$ vs. $V_{TG}$ and $V_b$. The model fits (purple and white lines) correspond to Landau level gaps at $\nu_T = \pm 2, \pm 6$ (top graphene layer) and at $\nu_B = \pm 2$ (bottom graphene layer), obtained using the effective junction bias voltage $\tilde{V}_J$ due to the nonlinear contact resistance (see SI Section 10). In this model calculation, we obtain $d_{TG} = 24$ nm (keeping $\varepsilon_{hBN} = 3.5\varepsilon_0$) and $d_{TB} = 2.3$ nm (using $\varepsilon_{WSe_2} = 7.6\varepsilon_0$). While these are the best fits of the model to the data, they do not fit perfectly due to the inhomogeneity of top gating across regions of the tip's contact area. These off-defect measurements average over variation in $C_{TG}$ across the entire tip area. In contrast, the defect-assisted tunneling measurements are sensitive to a specific region of the tip and consistently yield a match to the $d_{TG}$ fit (see Figs. 2g and 2h, and SI Section 12 with fits to different defects).

The above capacitances ($C_{TG}, C_{TB}, C_{BG}$) hold for all our measurements using different defects. The ratio of the defect couplings to the top and bottom layers, $C_T/C_B$, is naturally defect-dependent and is used to fit the on-defect tunneling data. Defects D1 and D2 have estimated $C_T/C_B$ ratios of 1/1.8 and 1.8/1, respectively.

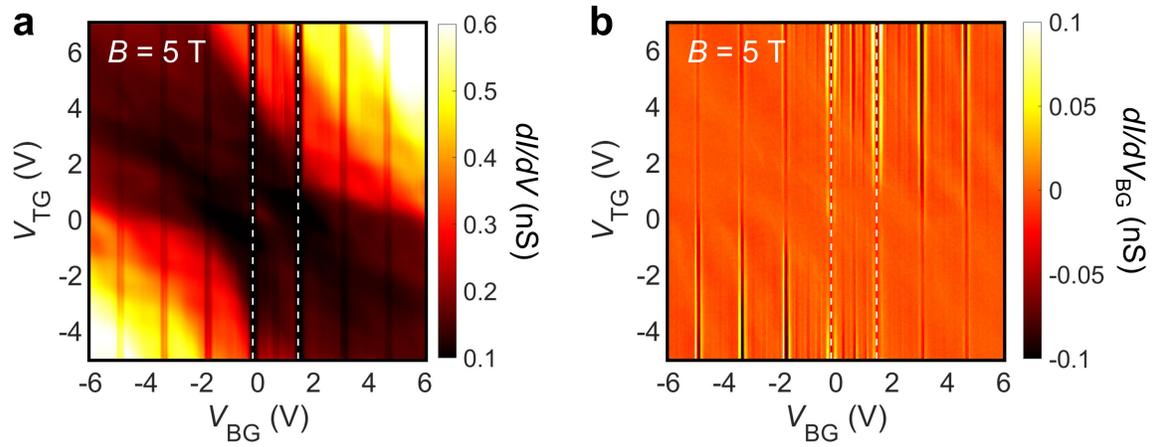

**Fig. S7: Calibrating the bottom gate geometric capacitance from Landau level measurements.** Measured off-defect conductance $dI/dV$ (**a**) and bottom gate transconductance $dI/dV_{BG}$ (**b**) vs. $V_{BG}$ and $V_{TG}$ at $B$ = 5 T. The white dashed lines mark the Landau level gaps at $\nu_B = \pm 2$ in the bottom graphene layer.

# SI Section 6. Spatial resolution of Atomic SET imaging

The spatial resolution of the defect imaging method is governed by the size of the defect's electronic wavefunction and the defect's standoff distance (SI Section 3) during scanning. In practice, the resolution can be directly obtained from the acquired images. Fig. S8a (reproduced from Fig. 1c in the main text) presents an image of the tunneling current, $I$, measured at a fixed bias. Multiple instances of the tip shape are visible, each corresponding to tunneling through a different defect. The resolution of the imaging is determined by the sharpness of the current drop at the boundary of these shapes, which corresponds to the point where a defect stops overlapping with the tip area. Fig. S8b displays the measured $I$ (black dots) along a linecut in the $x$ direction (white dotted line in Fig. S8a). The blue curve represents the best fit to the cumulative distribution function of a Lorentzian:

$$I = A \cdot \arctan\left(\frac{x-x_0}{\gamma}\right) + B \tag{S6.1}$$

The FWHM of this fit gives $2\gamma = 0.95$ nm, which demonstrates a spatial resolution of ~1 nm in defect imaging.

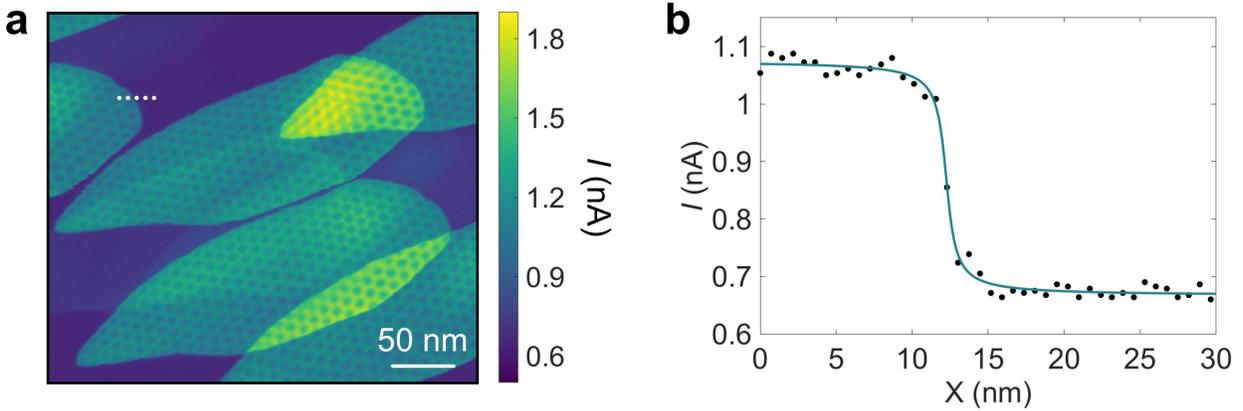

**Fig. S8: Determining the spatial resolution of defect imaging. a.** Map of $I(x,y)$ measured at a fixed bias $V_b = -0.7$ V at room temperature (reproduced from Fig. 1c of main text). **b.** $I$ (black dots) measured along a linecut in the $x$ direction (corresponding to white dashed line in **a**). This trace shows the drop-off in current as the tip is scanned to not overlap a defect. The blue line is a fit to equation S6.1 (see text).

# SI Section 7. Estimation of strain in the G/hBN moiré lattice on a QTM tip

In our fabrication process, prior to transfer onto a QTM tip, we fabricate a stack of graphene and hBN on a flat substrate and use Raman spectroscopy to confirm their close angle alignment. One might imagine that placing an aligned stack of G/hBN on the steep topography of the tip could destroy its moiré structure altogether, or at least introduce significant strain, particularly near the tip's apex. Below we analyze the detailed moiré structure imaged by the defect, demonstrating that the strain is rather small.

For the tip whose imaging is in Fig. S9, the heterostructure prior to transfer had a 2D peak broadening of $36 \pm 1$ cm$^{-1}$, corresponding to a moiré wavelength[3] of $11.9 \pm 0.3$ nm. This corresponds to a relative twist angle $\theta$ of 0.67° assuming a carbon-carbon bond distance $a = 0.142$ nm and a lattice mismatch $\varepsilon_{G,hBN} = 1.7\%$ between graphene and hBN[8]. We now wish to estimate the heterostrain of the moiré structure after it was transferred onto the QTM tip, which we measure directly from defect imaging of the tip (Fig. S9). Focusing on three representative moiré sites on the tip (white circles), we measure the distances to the six nearest neighbor moiré sites (in three directions) and list them in the table below:

| Site | $\lambda_1$ (nm) | $\lambda_2$ (nm) | $\lambda_3$ (nm) |
|---|---|---|---|
| 1 | 9.4 | 10.8 | 11.5 |
| 2 | 9.5 | 10.9 | 11.3 |
| 3 | 10.1 | 10.4 | 11.3 |

Each of these sites has an average moiré lattice constant of 10.6 nm (relative twist angle $\theta$ of 0.95°). However, for all three sites, there is a consistent variance in the moiré spacing in the three different directions. From this, we can estimate the heterostrain present in the moiré system with the following equation[8]:

$$\lambda = \frac{\sqrt{3}a}{\sqrt{(\varepsilon_{G,hBN}+\varepsilon_{strain})^2 + \theta^2}} = (10.6 - 312\varepsilon_{strain})\text{ nm} + O(\varepsilon_{strain}^2) \tag{S7.1}$$

Using $\theta = 0.95°$ and taking the extrema in measured lattice constants of 9.4 nm and 11.5 nm, we obtain a heterostrain $\Delta\varepsilon_{strain}$ of ±0.3%.

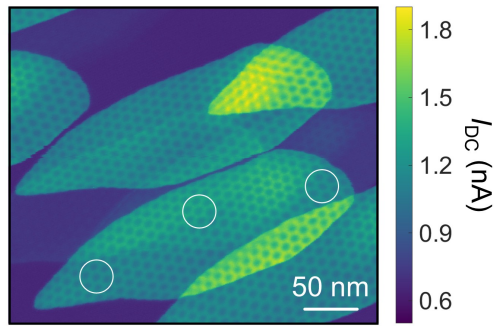

**Fig. S9: Estimation of strain from defect imaging of the G/hBN moiré-on-tip.** The white circles show three representative regions of the tip contact area that were used to calculate the superlattice periodicity in different directions (see table).

## SI Section 8. Potential sensitivity of the Atomic SET

In this section, we estimate the potential sensitivity of our Atomic SET technique. In practice, this sensitivity will depend on the exact gating and biasing conditions of the QD detector, and is optimized at the sides of the Coulomb blockade conductance peak where the transconductance $dI/dV_{BG}$ reaches its maximum.

To assess the potential sensitivity, we set the gate and bias voltages near the Coulomb blockade peak (using $V_b$ = 0.14 V, $V_{TG}$ = 1.4 V, and $V_{BG}$ = 0) of another low-energy defect (D3, not presented in the main text). A small AC voltage excitation is applied to the bottom gate and we measure $dI/dV_{BG}$. The sensitivity is derived from the signal-to-noise ratio obtained in this measurement. With a sampling time of 1 second and 3 mV AC rms excitation on the bottom gate, we achieve a signal-to-noise ratio (defined as mean divided by standard deviation) of 8.04 (Fig. S10). This gives us an equivalent bottom gate voltage noise of 380 µV/$\sqrt{Hz}$. The excitation applied to the bottom gate passes through the bottom graphene layer, which suppresses this excitation due to its finite compressibility. As a result, the potential modulation experienced by the QD is significantly reduced. To determine the potential sensitivity of the QD, we must therefore account for this suppression factor, which is calculated from the calibrated electrostatic model of the QTM junction that includes the geometric and quantum capacitances (see SI Sections 2 and 5). For the working point above, we calculate a suppression factor of approximately 60, yielding a potential sensitivity of $\delta\phi_D$ = 6 µV/$\sqrt{Hz}$ for our Atomic SET detector.

This extraordinary potential sensitivity is especially noteworthy when considering the expected trade-off between spatial resolution and potential sensitivity. Generally, detectors with higher spatial resolution exhibit lower potential sensitivity due to the inverse relationship between these two parameters. Specifically, the spatial resolution of a sensor is determined by its characteristic size $L$. Enhancing the spatial resolution by a factor of 2, by reducing this dimension to $L/2$, will reduce the sensor area by a factor of 4, thereby also reducing the capacitance between the sample and the sensor, $C$, by a factor of 4. The effective charge induced on the sensor by a small change in potential is $\delta n = C\delta\phi$. In state-of-the-art scanning SETs with strong transconductance amplification, charge noise is typically the dominant source of noise. Thus, improving spatial resolution by a factor of 2 would degrade potential sensitivity by a factor of 4.

Comparing to state-of-the-art nanotube-based SETs[9] which achieve a potential sensitivity of ~1 µV/$\sqrt{Hz}$ with a spatial resolution of 100 nm, we can appreciate the remarkable sensitivity of approximately 6 µV/$\sqrt{Hz}$ achieved with 1 nm spatial resolution by the Atomic SET.

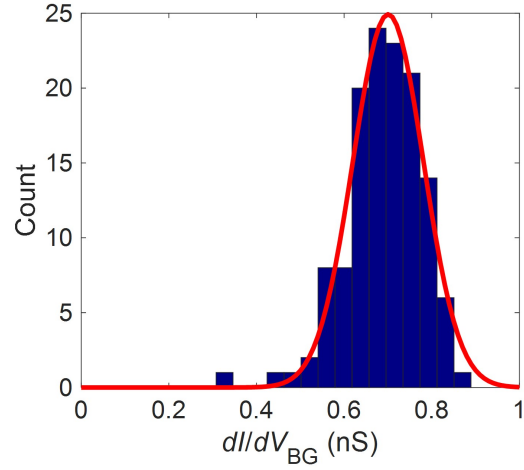

**Fig. S10: Determining the potential sensitivity of the Atomic SET.** The graph shows the histogram of measured transconductance $dI/dV_{BG}$ taken with 1 second integration time and an AC excitation of 3 mV on the bottom gate. The red curve is a Gaussian fit to the data, centered at 0.700 nS with standard deviation of $\sigma = 0.081$ nS. The potential sensitivity deduced from this measurement based on the junction parameters is 6 µV/$\sqrt{Hz}$ (see text).

## SI Section 9. Calculation of applied force and pressure in the experiment

The piezoresistive self-sensing readout of the cantilever deflection allows for direct calculation of the force applied to the QTM interface in the experiment. Fig. S11a shows the readout as a function of the $Z$ piezo motor position as the tip is retracted from the flat sensor device surface, where the initial point $\Delta Z = 0$ corresponds to the deflection setpoint maintained during all electrical measurements presented in the paper.

From $\Delta Z = 0$ to approximately $\Delta Z = -0.5$ μm, the cantilever is still in contact with the flat sample, with a constant deflection slope. The region of positive deflection $\Delta Z < -0.15$ μm results from the attractive vdW adhesion forces that oppose separation between the two sides of the interface as the motor is retracting. Around $\Delta Z = -0.5$ μm, the tip abruptly detaches from the sensor surface, after which the readout shows a steady value where the cantilever deflection is constant. Using the cantilever spring constant of 5 N/m, we can calculate the force $\Delta F = k \Delta Z$, where we take $\Delta Z$ to be the difference between the initial point and where the deflection readout matches the fully retracted value, highlighted in red. From this, we find that $\Delta F$ is approximately 500 nN. We can also calculate the vdW adhesion force in our junction by looking at the total $\Delta Z$ required to detach the tip from the sensor surface. From this, we obtain an attractive force of about 1500 nN.

Since our experiments directly image the contact area between the tip and sample, we can also determine the pressure applied to this surface. Fig. S11b shows an image of the tip contact area produced by probing the DC current through a single defect at a bias voltage of $-0.7$ V. The calculated area (white) is $5.3 \cdot 10^4$ nm$^2$. Thus, the pressure $P = F/A$ is approximately 10 MPa.

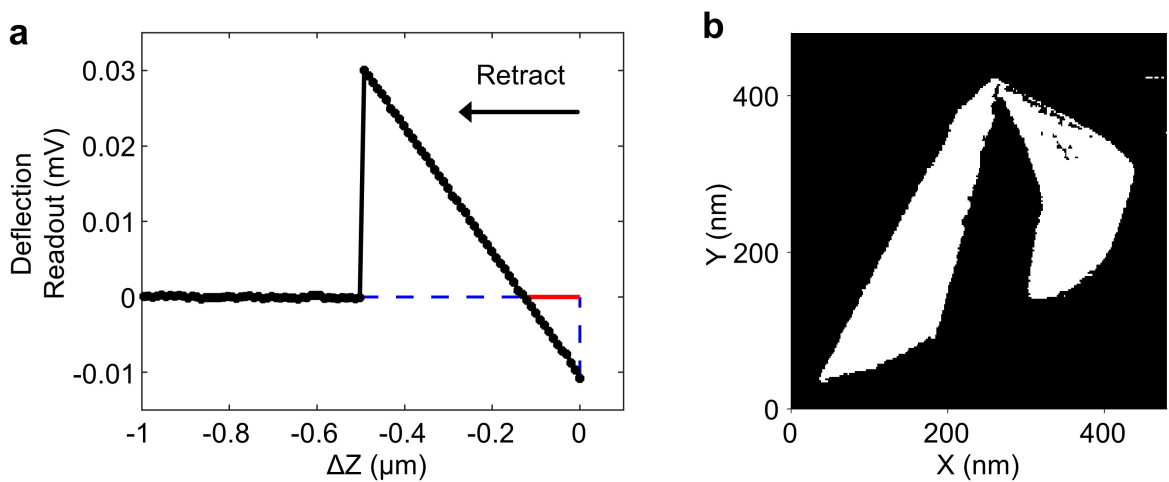

**Fig. S11: Determining the force and pressure applied to the interface. a.** Readout of cantilever deflection plotted as the cantilever is retracted from the flat surface by a distance $\Delta Z$. $\Delta Z = 0$ corresponds to the setpoint used during the experiment. The red line marks the distance over which a positive force is applied to the interface. **b.** Binary image of the DC current measured through a single defect at a bias of $-0.7$ V, used for calculating the contact area of the tip.

# SI Section 10. Nonlinear contact resistance model

Fig. 2g in the main text, reproduced here as Fig. S12a, shows the measured tunneling conductance through an atomic defect (D2) as a function of $V_{TG}$ and $V_b$. Strangely, at low bias, this measurement features a Coulomb blockade peak that is approximately independent of $V_b$ up to some threshold bias $V_{th}$, deviating from the typical Coulomb diamond structure observed in QDs. In a conventional QD at zero bias, a single conductance peak is observed as a function of gate voltage, corresponding to the resonance of the QD's energy level with the electrochemical potential of both the source and drain electrodes. When a finite bias voltage is applied, this peak splits into two distinct resonance conditions, appearing at two different gate voltages, $V_{TG}^{source}$ and $V_{TG}^{drain}$, where the QD energy level aligns with the electrochemical potential of either the source or the drain electrode. Moreover, there is a lower bound on the splitting between these two resonances given by $\Delta V_{TG}(V_b) \equiv V_{TG}^{source}(V_b) - V_{TG}^{drain}(V_b) = \alpha^{-1} V_b$ with $\alpha^{-1} > 1$, where $\alpha$ is the lever arm factor of the top gate.

Using additional transport data measured at finite frequencies, we demonstrate that this behavior is specific to the device and arises from its very high contact resistance until a threshold bias $V_{th}$, which then decreases sharply above this voltage. This suggests that up to $V_{th}$, the bias primarily drops across the contact rather than across the QTM junction itself, explaining why the measurements do not show bias dependence up to this voltage.

For this interpretation to be valid, the contact impedance much be greater that of the QTM junction below $V_{th}$. We confirm this in the DC limit, where the contact impedance dominates. However, as we increase the measurement frequency, the contact impedance continuously decreases until it becomes lower than the QTM junction impedance, at which point the data begin to exhibit a more conventional Coulomb diamond structure of a QD.

In Fig. S12b, we present a toy model that captures the key features of all measurements discussed below. In this model, the QTM junction is characterized by its resistance, $R_J$, which depends on the voltage across the junction, $V_J$, and the gate voltages. The capacitance of the QTM junction, $C_J$, is negligible within the frequency range explored and can be ignored. The QTM junction is in series with a resistance $R_c$, which is in parallel to a capacitance $C_c$. The resistance $R_c$ is highly nonlinear, behaving like a bidirectional diode: for bias voltages below $V_{th}$, $R_c \gg R_J$, but when the bias exceeds $V_{th}$, the diode "opens," and $R_c \ll R_J$. Additionally, we account for the capacitance between the lines in our fridge that are used in the experiment by $C_{line}$.

Although we cannot definitively determine the exact origin of $R_c$ and $C_c$, we believe they likely stem from the contact made to the graphene layer in the bottom device. As detailed in SI Section S1, this contact is formed by a graphite flake that electrically bridges a metallic electrode to the graphene layer, overlapping the latter with a finite area at a random relative twist angle. While such interfaces typically exhibit low resistance at small twist angles, we have recently shown[10] that at a large twist angle and at low temperatures ($T = 4$ K), the resistance of the interface can be extremely high – at $\theta = 30º$ it is nearly six orders of magnitude greater than at small angles. We have also demonstrated that this interface displays strong bidirectional diode-like behavior due to the activation of inelastic phonon emission channels at biases of a few tens of mV. These characteristics are consistent with the values observed in our measurements below.

To determine the capacitive elements $C_{line}$ and $C_c$, we performed several experiments measuring the frequency dependence of the total conductance under different conditions. In all cases, we applied zero DC bias and a small AC excitation on the bias voltage with $V_{rms} < 10$ mV (Fig. S12c).

First, we measured the conductance when the tip was fully retracted from the sensor surface (black points), which allowed us to determine the capacitance between the lines in our fridge, yielding $C_{line} = 22$ fF. Next, with the QTM tip in contact with the bottom sensor device, we adjusted $V_{TG}$ to place the QD in its zero-bias resonance condition and measured the frequency dependence of the conductance (red points, labeled 'on resonance' in Fig. S12c). At frequencies above approximately 10 Hz, the conductance showed a linear dependence on frequency, indicating a dominant capacitive contribution. From the linear frequency dependence, we extracted an effective capacitance of $C_{eff} = 0.3$ pF. At lower frequencies (< 10 Hz), the conductance saturated, revealing a low-frequency, low-bias contact resistance of approximately 10 GΩ.

Since the measured capacitance is more than two orders of magnitude greater than the geometric capacitance of the QTM junction ($C_J = C_{TB}A \approx 1.6$ fF, see SI Sections 2 and 5), it is more likely that it stems from a capacitive contribution at the contact, namely $C_c = C_{eff} = 0.3$ pF. A twisted contact interface between a graphite flake and graphene can exhibit high resistance while also providing significant shunt capacitance due to the geometric capacitance of the interface. Assuming a typical vdW interlayer spacing of ~0.33 nm, the measured $C_c$ would correspond to an area of approximately 12 μm², consistent with the typical overlap area between graphite and graphene in our devices.

Finally, we measured the junction conductance when the QD was not in resonance (blue points, labeled 'off resonance' in Fig. S12c). At moderate frequencies (> 10 Hz), the conductance was dominated by $C_{line}$. However, at low frequencies (< 10 Hz), the conductance was determined by the contact resistance, which was similar to that observed in the 'on resonance' measurements.

This model predicts that if the bias voltage, typically applied in the DC limit, is instead applied as an AC voltage at a moderate frequency, then the vertical line seen in Fig. S12a would transform into a more conventional looking Coulomb diamond that exhibits finite conductance over a range of gate voltages that expands with increasing bias voltage. Fig. S12d presents a conductance measurement as a function of AC bias ($V_{rms}$) and $V_{TG}$, taken at 2100 Hz with zero DC bias. Visibly, when using only an AC bias, we recover the expected behavior for QD-assisted tunneling, where the low-bias vertical line disappears and instead the window of finite conductance vs. $V_{TG}$ expands smoothly with increasing $V_{rms}$.

To illustrate the evolution of this nonlinear contact resistance from finite frequencies to the DC limit, we present a series of measurements similar to Fig. S12d taken at several frequencies (Fig. S12e). At the highest frequency of 2000 Hz (leftmost panel), no anomalous behavior is observed at low $V_{rms}$. However, as the frequency is decreased, a gap opens at low $V_{rms}$, resulting in a narrow window in $V_{TG}$ where there is a finite conductance. It is important to note that unlike in Fig. S12a where the y-axis ($V_b$) represents a fixed DC bias applied to the system, the y-axis in Fig. S12e corresponds to an AC bias $V_{rms}$. Namely, here the bias oscillates between $-\sqrt{2}V_{rms}$ and $+\sqrt{2}V_{rms}$, so direct comparison between the two measurements beyond the general features highlighted above is limited.

With the above model and supporting data, we can now return to examining the conductance as a function of DC bias shown in Fig. S12a. Below an applied DC bias of $|V_b| < V_{th}$, the system behaves as if there is no bias dropping on the QTM junction. However, above this threshold bias, the opening of the Coulomb diamond of the QD, i.e., $\Delta V_{TG}^{res}(V_b)$, behaves as expected. Therefore, we treat the nonlinear contribution of the contact element in series to the QTM junction as an ideal Zener diode with a threshold of 65 mV, resulting in an effective junction bias $\tilde{V}_J$:

$$\tilde{V}_J = \begin{cases} V_b - 65 \text{ mV}, & V_b > 65 \text{ mV} \\ 0, & |V_b| < 65 \text{ mV} \\ V_b + 65 \text{ mV}, & V_b < -65 \text{ mV} \end{cases} \tag{S10.1}$$

All electrostatic model fits to defect behavior use $\tilde{V}_J$ for the bias voltage in the junction.

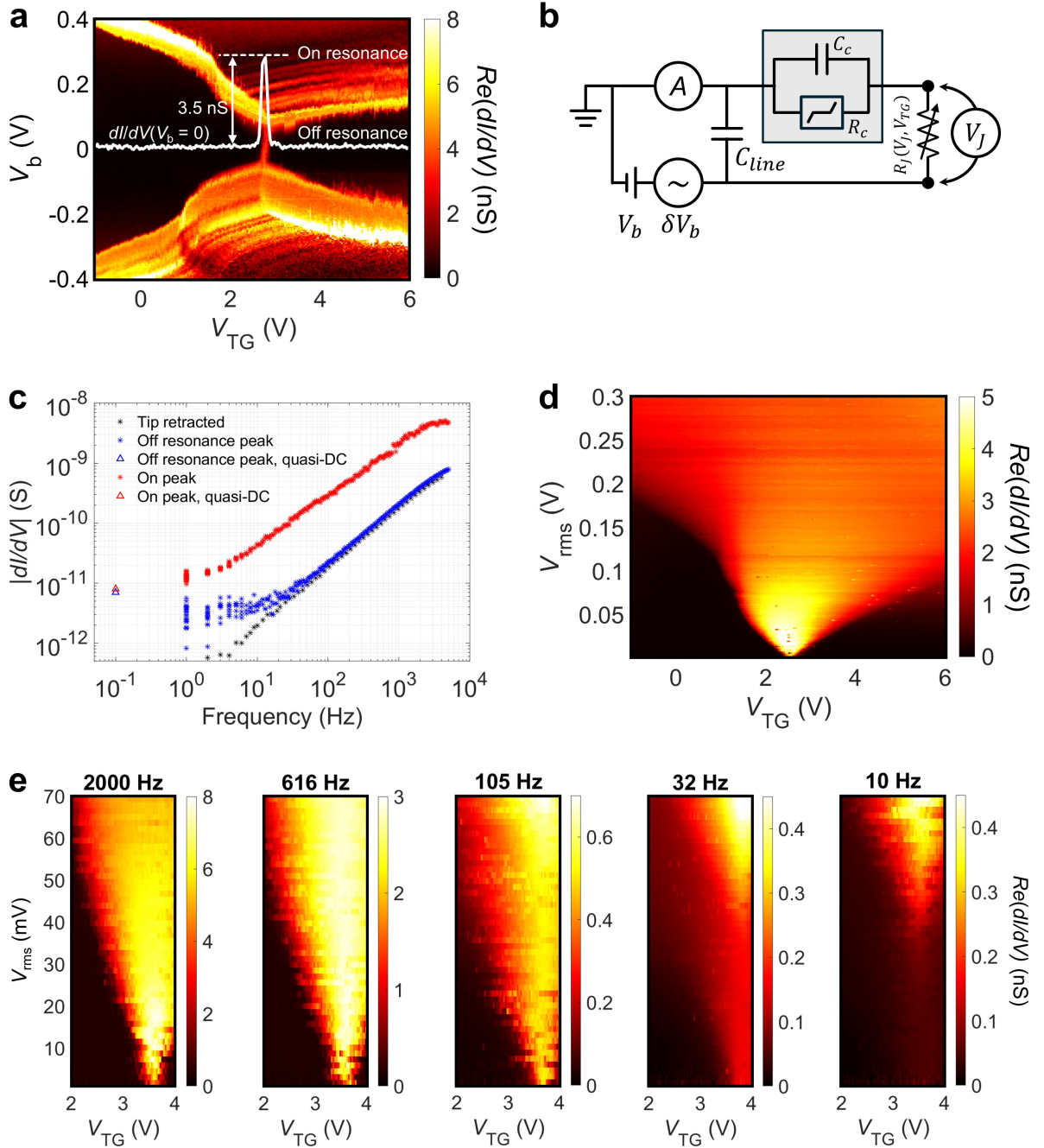

**Fig. S12: Nonlinear contact resistance model and additional supporting measurements. a.** Tunneling conductance through a single atomic defect (D2) vs. $V_{TG}$ and $V_b$, reproduced from Fig. 2g in the main text. In this measurement, we apply a large DC bias ($V_b$) and a small AC excitation of 1 mV at a frequency of 2100 Hz to determine $dI/dV$. The inset shows the linecut of $dI/dV$ at $V_b = 0$. **b.** Effective circuit diagram that accounts for the nonlinear contact resistance. The junction experiences an effective bias voltage $V_J$ (see text). The contact in our model (grey box) is comprised of a Zener diode in parallel to a capacitor, **c.** Frequency dependence of $|dI/dV|$ under different conditions: when the tip is retracted (black), when the tip in contact and overlapping a defect but $V_{TG}$ is tuned to be off of the resonance peak (blue), and when $V_{TG}$ is tuned to be on the resonance peak (red). **d.** Tunneling conductance similar to **a** but as a function of $V_{TG}$ and AC bias $V_{rms}$ at 2100 Hz. **e.** Frequency dependence of the low-AC bias data in **d**. As the frequency is decreased (left to right), the low-$V_{rms}$ signal gradually disappears due to the nonlinear contact resistance in series to the QTM junction.

# SI Section 11. Switching between different metastable defect states

Atomic defects in vdW materials have different energy levels as well as different charge states and local symmetry[11–13]. In our measurements, we observe that a defect can stay in a particular state for up to many weeks. However, we occasionally observe an abrupt switch between two or more metastable states of a defect. Fig. S13 shows three such examples in defects (D1, D2, and D3) located at different locations along the bilayer $WSe_2$. All scans were performed with $x$ as the fast scan axis and $y$ as the slow scan axis. They were taken at a fixed DC bias $V_b$ of –0.7 V and an AC excitation of 20 mV.

In all instances, we can see that at a certain point in the scan, there is an abrupt change in the values of both the DC current $I$ (top row) and the differential conductance $dI/dV$ (bottom row). It is highly likely that this change arises from a change in the state of a given defect, and not from switching to a different defect. We can rule this out since we can clearly see that the outline of the imaged tip shape remains unshifted after the abrupt change (to within our spatial resolution ~1 nm), and the likelihood of finding two relevant defects at this small separation is practically zero. Furthermore, the off-defect background tunneling signal is unaffected, confirming that the origin is not from the tip side of the interface. We see a clear anti-correlation between $I$ and $dI/dV$, which indicates a change in the defect energy, as opposed to a change in the tunneling rate $\Gamma$. A smaller defect energy results in more available inelastic tunneling channels, seen as an increase in $I$. In contrast, $dI/dV$ measures a residual change at $V_b$ further away from the elastic tunneling condition (*i.e.*, the QD resonance condition), resulting in a decrease in $dI/dV$.

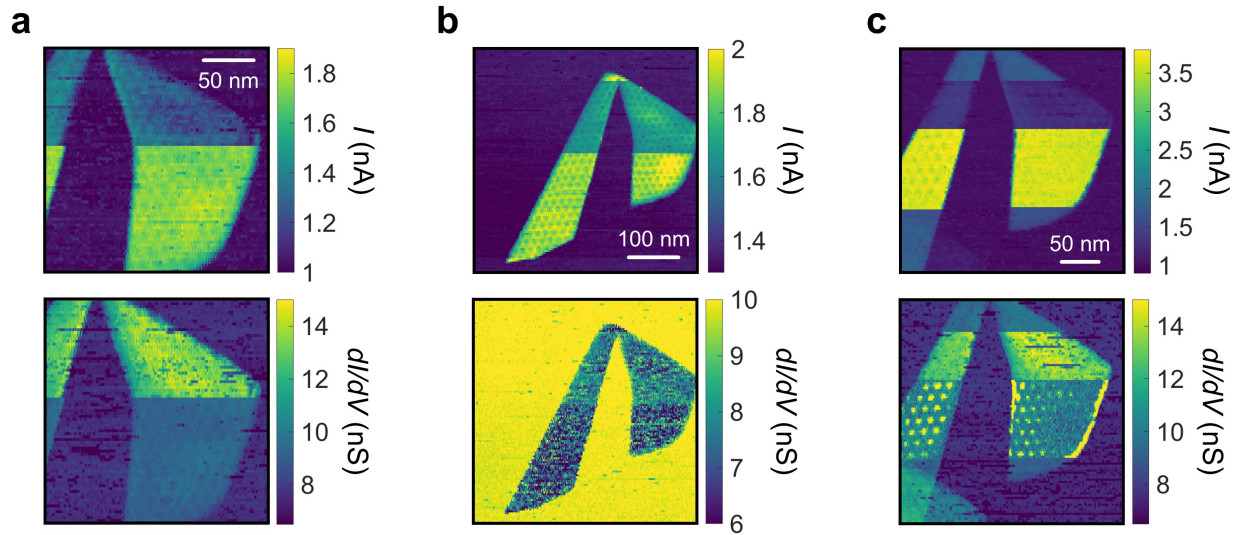

**Fig. S13: Imaging switching between metastable defect states. a-c.** Three examples of DC current $I$ and differential conductance $dI/dV$ maps measured at $V_b$ = –0.7 V with an AC excitation of 20 mV with three different defects D1, D2, and D3 (at different positions along the sensor layer). Simultaneous abrupt changes in $I$ (top) and $dI/dV$ (bottom) without any shift in the outline of the tip shape indicate the switching of the defect between metastable states.

## SI Section 12. Additional data: fits to different defects

In this section we present additional data measured on one of the defects in the main text (D1) and on another defect that does not appear in the main text (D4). We fit these data with the same electrostatic model used to fit defect D2 in Fig. 2 of the main text. Crucially, in all these fits, we keep the same values for all the geometric capacitances of the QTM junction ($C_{TG}, C_{BG}, C_{TB}$) and only fit the defect-specific parameters – the coupling ratio $C_T/C_B$ and the defect energy $E_D$. Similar to the analysis we did in Fig. 2, here the model also accounts for voltage drop occurring at the contacts, outside of the QTM junction, for $|V_b| < 65$ mV (SI Section 10). Fig. S14a shows the conductance ($dI/dV$) measured through defect D1 vs. $V_{BG}$ and $V_b$. The dashed lines are fits to the electrostatic model using $C_T/C_B = 1/1.8$ and $E_D = -96$ meV. Fig. S14b looks at this same defect (D1) in an applied magnetic field of 5 Tesla with model fits using the same ratio $C_T/C_B = 1/1.8$ and a rather close $E_D = -100$ meV. This change in $E_D$ arises from a shifted potential $\phi_D$ at different locations on the moiré superlattice on the tip. Figs. S14c and S14d are both obtained using another defect (D4), where Fig. S14c varies $V_{BG}$ at a fixed $V_{TG} = 7$ V and Fig. S14d varies $V_{TG}$ at a fixed $V_{BG} = 0$. In both cases, $C_T/C_B = 1.8$ and the defect energies are –235 meV and –251 meV, respectively. The different $C_T/C_B$ ratios for D1 and D4 attest to different relative positions of the defects to the top and bottom electrodes. The reasonable fits to these different defects ensure that the potential maps measured by the defects in the main text are well calibrated and are independent of the specific defect that is performing the measurement or its relative position to the top and bottom layers.

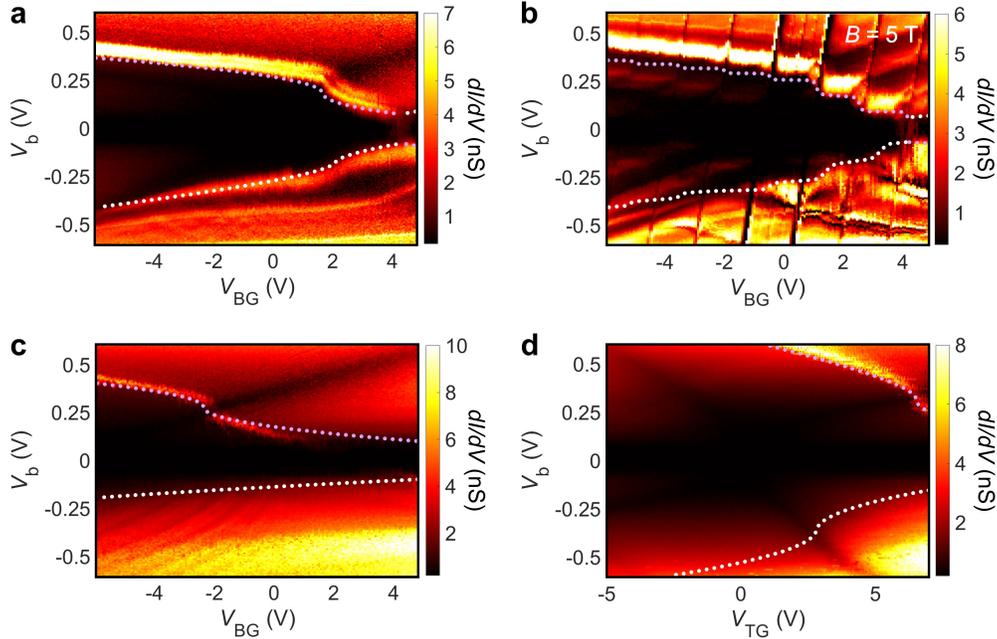

**Fig. S14. Measurements and fits to other defects. a.** Conductance $dI/dV$ vs. bottom gate $V_{BG}$ and bias $V_b$ for a low-energy defect (D1). The fits (dashed lines) follow the electrostatic model described in SI Section 2. **b.** $dI/dV$ vs. $V_{BG}$ and $V_b$ for the same defect (D1) taken at $B = 5$ T. The steps in the data and in the fits (dashed lines) occur due to Landau quantization in the top and bottom graphene layers. **c.** $dI/dV$ vs. $V_{BG}$ and $V_b$ at a fixed $V_{TG} = 7$ V for another defect (D4) with model fits (dashed lines). **d.** Same as in **c** but with $V_{BG} = 0$ V.

## SI Section 13. Comparing moiré potential mapping using top and bottom gates

In the experiment, as the defect scans across the moiré, it gets gated by the local electrostatic potential, $\phi_m^D(x,y)$. We use a gate voltage to bring the Coulomb blockade peak of the defect back into resonance. In the figures presented in the main text, this was achieved using the top gate voltage, $V_{TG}$. However, since our setup also includes a bottom gate, we can perform a similar scan where the Coulomb peak is brought back into resonance using the bottom gate voltage, $V_{BG}$. Given that $V_{TG}$ and $V_{BG}$ affect the charge densities of the top and bottom graphene layers differently, comparing these two sets of measurements serves as an important consistency check for our technique.

Fig. S15 plots the $dI/dV$ measured at zero bias as a function of $V_{BG}$ and $V_{TG}$, illustrating how the zero-bias Coulomb peak evolves in the $V_{BG}$-$V_{TG}$ plane. We selected a point along this curve, which determines the carrier densities in the top and bottom layers, and performed two types of measurements: a $dI/dV(x,y,V_{TG})$ map where $V_{TG}$ was scanned (following the vertical dashed arrow) to track the Coulomb blockade peak at each spatial location, and a $dI/dV(x,y,V_{BG})$ map where $V_{BG}$ was scanned instead (following the horizontal dashed arrow). Both measurements were conducted at zero DC bias with an AC excitation of 10 mV.

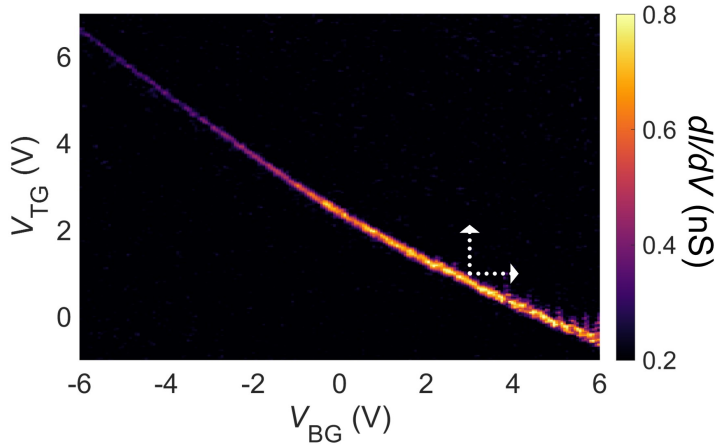

**Fig. S15: Zero-bias $dI/dV$ vs. $V_{BG}$ and $V_{TG}$.** On-defect conductance taken at the moiré site corresponding to highest electrostatic potential using defect D1. The white dashed arrows show the directions along which the potential scans in Fig. S16a and S16b were performed.

Figs. S16a and S16b show the potential maps extracted from these two independent scans, revealing a striking similarity between the measurements. To quantify this further, Fig. S16c plots the difference between the two maps, which amounts to only a few percent of the measured values. This small discrepancy clearly demonstrates the internal consistency of our measurement technique and analysis procedures. The map in Fig. S16a appears as the middle panel in Fig. 3d of the main text (corresponding to filling of the moiré-on-tip of $\nu = 1.3 \pm 0.2$).

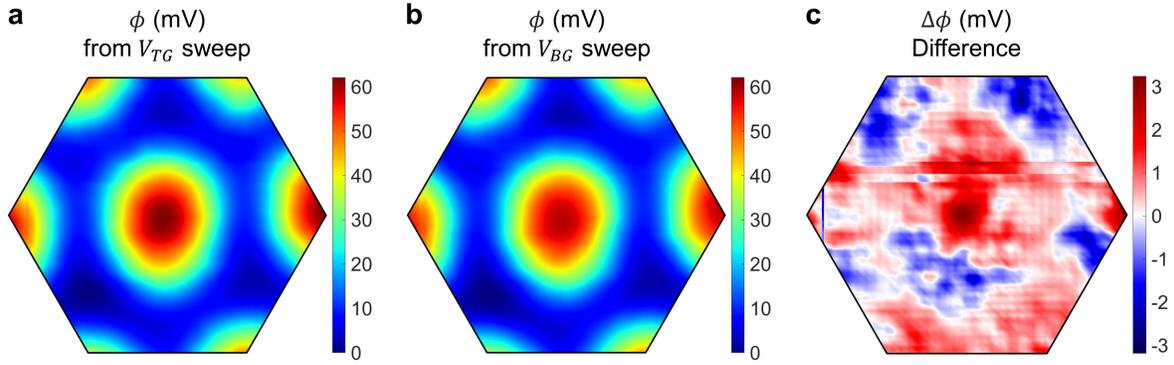

**Fig. S16: Potential maps using different scanned gate voltages. a.** Averaged potential map $\phi(x,y)$ obtained by sweeping $V_{TG}$ at a fixed $V_{BG}$ = 3 V. **b.** $\phi(x,y)$ averaged over the same moiré sites as in **a**, obtained by sweeping $V_{BG}$ at a fixed $V_{TG}$ = 1 V. **c.** The difference $\Delta\phi(x,y)$ between the potential maps in **a** and **b**, which is only a few percent of the measured potential, demonstrating the internal consistency of our measurement technique and analysis procedures.

# SI Section 14. Calculation of the self-consistent Hartree potential

When graphene is stacked on hBN, the lattice mismatch between the two hexagonal lattices generates a moiré potential. Following the notation of Ref. 8, the effective non-interacting Hamiltonian for this heterostructure is expressed as $\mathcal{H} = \hbar v_F \mathbf{p} \cdot \boldsymbol{\sigma}\tau_0 + H_0(\mathbf{r})\sigma_0\tau_0 + H_z(\mathbf{r})\sigma_3\tau_3 + \mathbf{H_{xy}}(\mathbf{r}) \cdot \boldsymbol{\sigma}\tau_3$, where $v_F$ is the monolayer graphene Fermi velocity and the sublattice and valley pseudospin degrees of freedom are encoded through the Pauli matrices $\boldsymbol{\sigma}$ and $\boldsymbol{\tau}$ that act on the four-component wavefunction $(\Psi_{A,K}, \Psi_{B,K}, \Psi_{B,K'}, -\Psi_{A,K'})^T$.

The parameters for this effective Hamiltonian are calculated in Ref. 8 using a local stacking approximation, where the stacking energies were obtained from *ab initio* calculations under different relaxation cases. The most relevant case for the present study of encapsulated graphene is referred to as "Relaxed-XY $\beta$," in which only in-plane relaxation was allowed, while the interlayer separation was kept constant. This scheme accounts for the effects of relaxation on the interlayer couplings, intralayer hoppings, and the pseudomagnetic fields induced by lattice relaxation. The parameterization is expressed as:

$$H_0(\mathbf{r}) = \alpha_0 \sum_{m=1}^{6} e^{i\mathbf{G_m}\cdot\mathbf{r}} + \beta_0 \sum_{m=1}^{6} i(-1)^{m-1} e^{i\mathbf{G_m}\cdot\mathbf{r}} \tag{S14.1}$$

$$\mathbf{H_{xy}}(\mathbf{r}) = \alpha_{xy} \sum_{m=1}^{6} i(-1)^{m-1} (\hat{\mathbf{z}} \times \widehat{\mathbf{G}}_\mathbf{m}) e^{i\mathbf{G_m}\cdot\mathbf{r}} + \beta_{xy} \sum_{m=1}^{6} i(\hat{\mathbf{z}} \times \widehat{\mathbf{G}}_\mathbf{m}) e^{i\mathbf{G_m}\cdot\mathbf{r}} \tag{S14.2}$$

$$H_z(\mathbf{r}) = \alpha_z \sum_{m=1}^{6} i(-1)^{m-1} e^{i\mathbf{G_m}\cdot\mathbf{r}} + \beta_z \sum_{m=1}^{6} e^{i\mathbf{G_m}\cdot\mathbf{r}} \tag{S14.3}$$

where we take the parameters $(\alpha_0, \beta_0, \alpha_{xy}, \beta_{xy}, \alpha_z, \beta_z) = (-3.22, -5.28, -13.59, 2.60, -6.72, 2.89)$ meV from Ref. 8. Here, $\mathbf{G_m}$ are the first shell of the moiré reciprocal lattice vectors, $\widehat{\mathbf{G}}_\mathbf{m}$ represent their corresponding unit vectors, and $\hat{\mathbf{z}}$ is a unit vector in the out-of-plane direction. Following Ref. 8, we also use $v_F = 0.97 \times 10^6$ m/s. The coordinate system is chosen such that $\mathbf{G_m} = |G_0|\mathbf{R}\left[\frac{(m-1)\pi}{3}\right](\cos(\pi/6), \sin(\pi/6))$, where $\mathbf{R}[\phi]$ is a 2D rotation matrix by angle $\phi$, $|G_0| = 4\pi/\sqrt{3}L$, and $L$ is the moiré period.

New in this work is the calculation of the self-consistent Hartree potential that arises from the variation of the charge density with the same periodicity as the moiré potential. We account for the self-consistent screening response of the electrons in the graphene by including a Hartree potential, where the mean-field interacting Hamiltonian becomes[14–16]:

$$\mathcal{H}_{Hartree} = \mathcal{H} + V_H \tag{S14.4}$$

$$V_H = \int d^2\mathbf{r}\, \Psi^\dagger(\mathbf{r}) v_H(\mathbf{r}) \Psi(\mathbf{r}) \tag{S14.5}$$

$$v_H(\mathbf{r}) = \int d^2\mathbf{r}' \frac{e^2}{\epsilon|\mathbf{r}-\mathbf{r}'|}\rho(\mathbf{r}') \tag{S14.6}$$

Here, $e$ is the electric charge, $\epsilon$ is the dielectric constant, and $\rho(\mathbf{r})$ represents the fluctuations of the charge density with contributions from all occupied wavefunctions starting from the bottom of the bands. Since the Hartree potential is diagonal in the valley and sublattice basis, the Fourier transform yields:

$$v_H(\mathbf{k}) = \sum_{\mathbf{G}\neq 0} \frac{e^2}{\epsilon|\mathbf{G}|}\rho_\mathbf{G}\Psi^\dagger(\mathbf{k})\Psi(\mathbf{k}-\mathbf{G}) \tag{S14.7}$$

where $\rho_\mathbf{G}$ are the Fourier components of the density matrix, given by:

$$\rho_\mathbf{G} = 4 \sum_{n,\mathbf{k}\in occ}\sum_{\xi,\mathbf{G}'} \phi_{\mathbf{G}'}^{\xi,n*}(\mathbf{k})\phi_{\mathbf{G}+\mathbf{G}'}^{\xi,n}(\mathbf{k}) \tag{S14.8}$$

Here, $n$ represent the band indices, $\xi$ denote the sublattice indices, and the summation is performed over all occupied states. The factor of 4 accounts for spin and valley degeneracy. In momentum space, the Hartree potential couples the state $\Psi^\dagger(\mathbf{k})$ to the state $\Psi(\mathbf{k}-\mathbf{G})$ with a coupling strength of $\frac{e^2}{\epsilon|\mathbf{G}|}\rho_\mathbf{G}$.

The equations outlined above need to be solved self-consistently. Starting with the non-interacting $\rho_\mathbf{G}$ values, we proceed by diagonalizing the full Hamiltonian in equation S14.4 to obtain new eigenvalues and eigenvectors. The new density matrix is then computed using equation S14.8. These steps are iterated until a self-consistent solution is found. However, we found that direct iteration without mixing often leads to instability or divergence, and thus a mixing scheme was necessary to achieve convergence. In this calculation, we adopt $\epsilon = 4$ to account for the background dielectric constant of hBN. Fig. S17 illustrates both the one-shot (setting $V_H = 0$) and self-consistent Hartree potentials based on $H_0(\mathbf{r})$ along a high-symmetry line in real space. These were calculated at charge neutrality, where the individual Hartree responses to $\mathbf{H}_{xy}$ and $H_z$ vanish by symmetry of the wavefunctions (Fig. S18). The Hartree potential preserves the symmetry of $H_0(\mathbf{r})$, exhibiting maxima in regions where the corresponding pseudopotential is at a minimum. This occurs because the Hartree potential reflects the electron density distribution in real space. A minimum in $H_0$ will attract carriers, and this increased carrier density will in turn give a larger value for the Hartree potential. Additionally, the self-consistent calculation reduces the magnitude of the one-shot Hartree potential by approximately half.

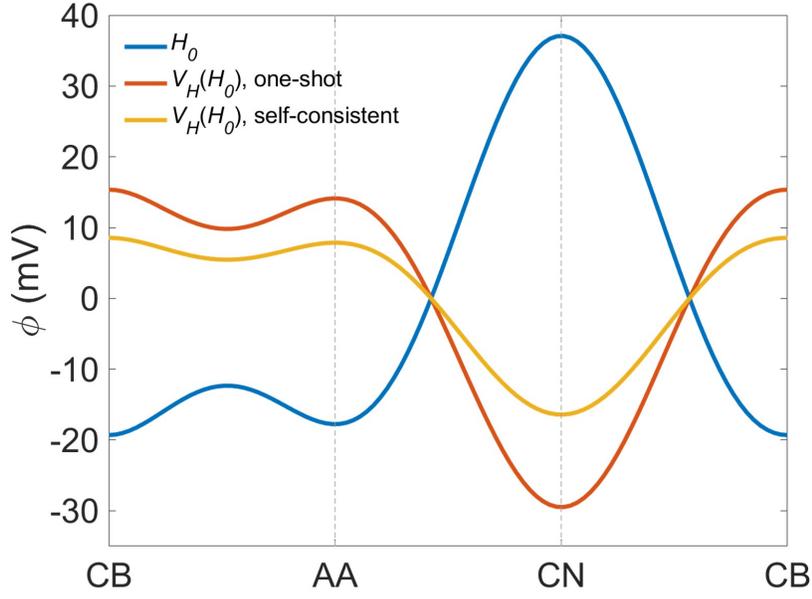

**Fig. S17: Calculated Hartree potential.** Calculations of the one-shot and self-consistent Hartree potentials for G/hBN considering only the pseudoelectric potential $H_0(\mathbf{r})$ from equation S14.1 at charge neutrality. The Hartree potential mirrors the symmetry of $H_0(\mathbf{r})$ but with opposite sign. Self-consistency reduces the magnitude by about half.

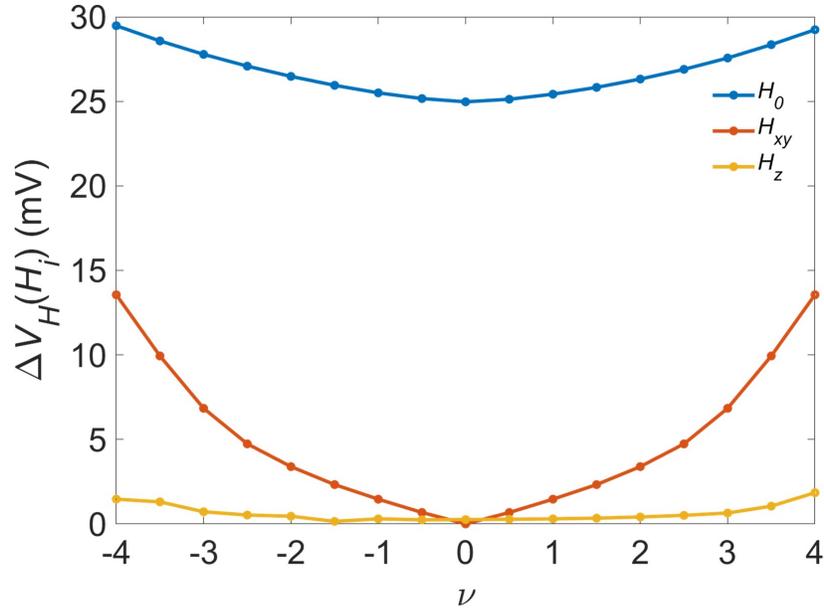

**Fig. S18: Density dependence of the Hartree potential.** Self-consistent Hartree potentials $\Delta V_H(H_i)$ calculated independently for the three terms $H_0$, $H_{xy}$, and $H_z$ as a function of moiré filling factor $\nu$. For each calculation, the parameters $\alpha_i$ and $\beta_i$ used to define the effective Hamiltonian (equations S14.1-3) were unchanged from the stated values for $H_i$, and set to 0 for the other two terms. The individual Hartree potentials of $H_{xy}$ and $H_z$ vanish at charge neutrality.